\documentclass[12pt]{iopart}
\usepackage{amssymb}
\usepackage{amsfonts}
\usepackage{graphicx}
\usepackage{amsthm}

\newtheorem{theorem}{Theorem}
\newtheorem{proposition}{Proposition}
\newtheorem{lemma}{Lemma}

\newtheorem{definition}{Definition}

\newtheorem{remark}{Remark}
\newtheorem{example}{Example}

\begin{document} 
\title[Role of a Phase Factor in the Boundary Condition of a One-Dimensional Junction]
	{Role of a Phase Factor in the Boundary Condition of a One-Dimensional Junction}
 \author{Yoshiyuki Furuhashi$^{1}$, Masao Hirokawa$^{1,\ast}$, Kazumitsu Nakahara$^{1}$, and Yutaka Shikano$^{2,3,\dagger}$}
 \address{$^1$ Department of Mathematics, Okayama University, 3-1-1 Tsushima-Naka, Kita, Okayama 700-8530, Japan}
 \address{$^2$ Department of Physics, Tokyo Institute of Technology, 2-12-1 Oh-Okayama, Meguro, Tokyo 152-8551, Japan}
 \address{$^3$ Department of Mechanical Engineering, Massachusetts Institute of Technology, 77 Massachusetts Avenue, Cambridge, MA 02139, USA}
 \ead{$^\ast$\mailto{hirokawa@math.okayama-u.ac.jp}, $^\dagger$\mailto{shikano@mit.edu}}
 \date{\today}
\begin{abstract}
	One-dimensional quantum systems can be experimentally studied in recent nano-technology like the carbon nanotube and the nanowire. 
	We have considered the mathematical model of the one-dimensional Schr\"{o}dinger particle with a junction and have analyzed the phase 
	factor in the boundary condition of the junction. We have shown that the phase factor in the tunneling case appears  
	in the situation of the non-adiabatic transition with the three energy levels in the exact WKB analysis.
\end{abstract}
\submitto{\JPA}
\maketitle

\section{Introduction} \label{section:intro} 
Quantum phases take the crucial role of the quantum interference 
and coherence. It is well known that the global phase factor for the 
quantum state is undetectable. 
On the other hand, since the relative phase is detectable by interference patterns such as 
the Young double-slit experiment for the electron~\cite{Tonomura}
and the molecule~\cite{Arndt}, this quantity 
is meaningful. By the experimental demonstration of the 
delayed choice experiment or the quantum eraser~\cite{Choi}, 
we know that this quantity depends on the operational set-up. 
Aharonov and Bohm predicted a phase of the electrically charged particle from an electromagnetic 
potential~\cite{Aharonov}, which is known as the Aharonov-Bohm phase. 
Furthermore, Berry predicted a phase acquired over the course of a cycle with adiabatic processes 
resulting from the geometrical properties of the 
parameter space of the Hamiltonian~\cite{Berry}, which is known as the Berry phase 
or the geometric phase. The 
Aharonov-Bohm phase~\cite{AB1,AB2,AB3} and the Berry phase~\cite{BP1,BP2,BP3} are experimentally realized. They are given by the Hamiltonian 
decided from the operational set-up. However, there also exists a phase factor in the 
boundary condition, which is not decided by the Hamiltonian but is decided by the situation of a quantum particle. 
In this paper we address the latter quantum phase under a one-dimensional system like the following physical set-up.

Recent development on experimental techniques has provided a way to study the one-dimensional quantum physics, for instance, 
see Ref.~\cite{Giamarchi}. In this paper, we focus on the one-dimensional electron transport system. 
The electron on the single-wall carbon nanotube~\cite{Saito} and the nanowire made of semiconductor materials such as InP~\cite{Gudiksen}, 
InAs/InP~\cite{Bjork}, GaAs/GaP~\cite{Gudiksen}, and Si/SiGe~\cite{Wu} can be described as the one-dimensional quantum system. 
This can be controlled by the application of the technique on a single-electron transistor, which is a device in which electrons 
tunnel one at a time through a small island connected to two leads via a tunnel junction, in the single-wall carbon nanotube~\cite{Tans} 
and the InP nanowire~\cite{Franceschi}. Furthermore, two carbon nanotubes electrically can be connected via a junction such as a 
gold particle~\cite{Thelander}. See more examples on the connected carbon nanotubes in Ref.~\cite{Carbon}. 
We will consider a mathematical model of the one-dimensional quantum system with a junction throughout this paper.

As is well known, a physical observable is 
described by a self-adjoint operator~\cite{Neumann}. 
Thus, the set $D(H)$ of all wave functions 
of a Hamiltonian $H$ should be determined 
so that $H$ becomes a self-adjoint operator. 
Usually, we begin with considering the action of 
an energy operator $H_{0}$ for the Hamiltonian $H$ 
on a domain $D(H_{0})$ in which the energy operator 
$H_{0}$ is not self-adjoint since it is smaller 
than $D(H)$. 
Thus, we seek the Hamiltonian $H$ as an extension 
of $H_{0}$. 
This extension is called a \textit{self-adjoint extension}~\cite{rs2}.  
As the boundary condition for a physical set-up 
is fixed, a self-adjoint extension is determined 
so that the extension corresponds to the boundary 
condition. 
 
It is already known that a phase factor appears 
in a boundary condition for a self-adjoint extension 
of a momentum operator on a non-Euclidean space~\cite{rs2,hir00}. 
In the case of Hamiltonians, however, a phase factor does not always appear 
in the boundary condition, 
for instances, Example 2 in Ref. \cite[\S X.1]{rs2}, Theorem 3.1.1 in Ref.~\cite{AGH-KHP}, and Eq. (1.1) in Ref.~\cite{exner}. 
Thus, in this paper we make a realization of 
the above physical set-up to obtain a mathematical model 
to consider a Schr\"{o}dinger particle 
in a line with a junction. 
In our mathematical idealization, the junction is 
represented by the closed interval 
$[-\Lambda , \Lambda]$ and the Schr\"{o}dinger particle 
moves in $(-\infty , -\Lambda)\cup(\Lambda , \infty)$:  
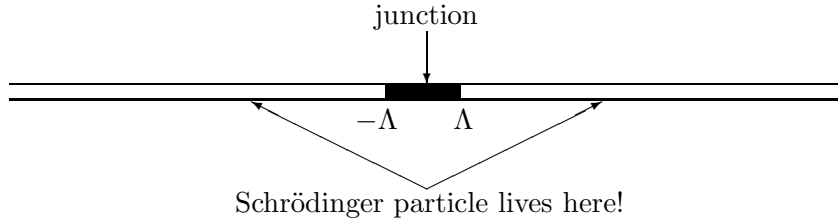
\begin{center}
   \setlength{\unitlength}{1mm}
   \begin{picture}(110,30)
      \thinlines
        \put(0,15){\line(1,0){50}}
        \put(0,17){\line(1,0){50}}
        \put(46,11){\small $-\Lambda$}
        \put(48.5,25){\small junction}
        \put(55.5,24){\vector(0,-1){7}}
        \put(55,16){\makebox(0,0){\rule{10mm}{2.2mm}}}
        \put(59,11){\small $\Lambda$}
        \put(55.5,3){\vector(-2,1){23.3}}
        \put(55.5,3){\vector(2,1){23.3}}
        \put(30,0){\small Schr\"{o}dinger particle lives here!}
        \put(60,15){\line(1,0){50}}
        \put(60,17){\line(1,0){50}}
   \end{picture}
\end{center}
For example, we can take a non-relativistic electron 
as the Schr\"{o}dinger particle~\footnote{In the case of carbon nanotubes, the non-relativistic electron 
can be taken as the excitation of the Tomonaga-Luttinger liquid~\cite{Carbon}.}, and then, 
the junction is made from an insulator. 
We investigate the phase factor determined by 
the boundary conditions at the two edges 
($x=-\Lambda$ and $x=\Lambda$) of the junction 
when the Schr\"{o}dinger particle tunnels through 
the junction.  
In the near future we will consider controlling 
the phase factor determined by the boundary conditions 
using the Aharonov-Bohm phase obtained by 
a magnetic field through the junction only. 

Our results characterize the boundary conditions 
for the point interaction  
given in Refs.~\cite{AGH-KHP} and \cite{eg} based on  
whether the Schr\"{o}dinger 
particle tunnels through the junction or not. 
More precisely, the boundary condition 
in the case where the Schr\"{o}dinger particle 
does not tunnel through the junction 
(as in Theorem \ref{theo:1}) corresponds to 
that for the point interaction given in Ref.~\cite{AGH-KHP} 
(see Remark \ref{rem:non-tunneling}). 
On the other hand, the boundary condition 
in the case where the Schr\"{o}dinger particle 
does tunnel through the junction 
(as in Theorems \ref{theo:4} and \ref{theo:2}) 
corresponds to that for the point interaction given 
in Ref.~\cite{eg} (see Remark \ref{rem:tunneling}). 
Namely, our results tell us that the generalized 
boundary condition given in the unfortunately unpublished 
paper~\cite{eg} is important in the light of 
the Schr\"{o}dinger particle tunneling the junction. 

Our paper is constructed as follows. 
In Sec.~\ref{section:pm} we recall some well-known 
facts and formulate our problem. 
In Sec.~\ref{section:pf} we investigate the boundary 
conditions of wave functions, dividing them into two cases. 
In the first case we handle the Schr\"{o}dinger particle 
not tunneling through the junction. 
In this case we can completely classify the type of boundary conditions 
of which type corresponds to that of in Ref.~\cite{AGH-KHP}. 
In other case we consider the Schr\"{o}dinger particle 
tunneling through the junction. We give another type of boundary condition which corresponds 
to that for the point interaction given in Ref.~\cite{eg}. Furthermore, 
this phase corresponds to one obtained by the exact WKB analysis in the model of the three-levels non-adiabatic 
transition inside the junction. Section \ref{section:conc} is devoted to the summary and the discussions.
\section{Preparations and Our Model} \label{section:pm}
\subsection{Mathematical Notations}
In this section, we prepare some mathematical terms and notions. 
For every operator $A$ acting in a Hilbert space, 
$D(A)$ expresses the set of all vectors on which the operator $A$ can act. 
For instance, $D(A)$ is the set of all wave functions 
as $A$ is an energy operator. 
$D(A)$ is called the \textit{domain} of the operator $A$. 
For operators $A$ and $B$ we say $A$ is equal to $B$, i.e., 
$A=B$ if and only if $D(A)=D(B)$ and $A\psi = B\psi$ for every 
$\psi \in D(A)=D(B)$, where $\psi \in D(A)$ means the vector 
$\psi$ belongs to the domain $D(A)$. 
When $D(A) \subset D(B)$ and  $A\psi = B\psi$ for every 
$\psi \in D(A)$, we say that the operator $B$ is 
an \textit{extension} of the operator $A$, and we express that 
by $A \subset B$. 
$D(A^{*})$ expresses the set of all vectors $\varphi$ 
satisfying the following for an operator $A$: 
there is a vector $\phi_{{}_{A}}$ 
so that $\langle\varphi|A\psi\rangle = 
\langle\phi_{{}_{A}}|\psi\rangle$ for every $\psi \in D(A)$.    
Then, the \textit{adjoint operator} $A^{*}$ of the operator $A$ 
is given by $A^{*}\varphi = \phi_{{}_{A}}$ for every 
$\varphi \in D(A^{*})$. 
Note that the domain $D(A^{*})$ has to be dense 
in the Hilbert space since the adjoint operator 
$A^{*}$ is determined uniquely. 
The operator $A$ is said to be \textit{symmetric} 
as $A\subset A^{*}$, 
and moreover, the operator $A$ \textit{self-adjoint} 
if and only if $A=A^{*}$. 
Thus, when an operator $B$ is called 
a \textit{self-adjoint extension} 
of an operator $A$, the operator $B$ satisfies 
$B=B^{*}$ and $A \subset B$. 
$D(\overline{\!A})$ expresses the set of all vectors $\psi$ 
satisfying the following conditions for an operator $A$: 
there is a sequence $\left\{\psi_{n}\right\}_{n}$ 
of vectors $\psi_{n}\in D(A)$ so that 
sequences $\left\{\psi_{n}\right\}_{n}$ and 
$\left\{A\psi_{n}\right\}_{n}$ converge, 
and $\psi = \lim_{n\to\infty}\psi_{n}$. 
Then, the \textit{closure} $\overline{\!A}$ of 
the operator $A$ is defined by 
$\overline{\!A}\psi := \lim_{n\to\infty}A\psi_{n}$. 
We say that the operator $A$ is \textit{closed} 
if $A=\overline{\!A}$. 
It is well known that a self-adjoint operator is closed. 

Following Ref.~\cite[Example 2 in \S X.1]{rs2}, 
we recapitulate some facts on self-adjoint extension here. 
For the subset $\Omega$ of the line $\mathbb{R}:=
(-\infty , \infty)$ 
$C_{0}^{\infty}(\Omega)$ expresses the set of all 
infinitely differentiable functions on $\Omega$ 
with their individual compact supports in $\Omega$. 
Here the support of a function $\psi$ on $\Omega$ 
is the closure $\overline{\left\{ x\in\Omega\, |\, 
\psi(x)\ne 0\right\}}$ of the set 
$\left\{ x\in\Omega\, |\, 
\psi(x)\ne 0\right\}$. 
$AC^{2}(\Omega)$ expresses the set of all 
absolutely continuous functions $\psi$ on $\Omega$ 
so that $\psi'$ is also absolutely continuous and 
$\psi''$ is square integrable on $\Omega$. 
It should be noted that the Lebesgue theorem states that 
absolutely continuous function $\psi$ has its 
differentiable $\psi'$  almost everywhere. 
 
The regions $(-\infty , -\Lambda)$ 
and $(\Lambda , \infty)$ is denoted as $\Omega_{\mathrm{L}}$ 
and $\Omega_{\mathrm{R}}$ for an arbitrarily fixed 
constant $\Lambda>0$, respectively. 
We define energy operators $H_{\mathrm{L}00}$ and 
$H_{\mathrm{R}00}$ 
by 
$H_{\mathrm{L}00}:=- d^{2}/dx^{2}$ with $D(H_{\mathrm{L}00}):= 
C_{0}^{\infty}(\Omega_{\mathrm{L}})$ 
and 
$H_{\mathrm{R}00}:=- d^{2}/dx^{2}$ with $D(H_{\mathrm{R}00}):= 
C_{0}^{\infty}(\Omega_{\mathrm{R}})$ 
respectively. 
Set $H_{\mathrm{L}0}$ and $H_{\mathrm{R}0}$ as 
$H_{\mathrm{L}0}:=\overline{H_{\mathrm{L}00}}$ 
and $H_{\mathrm{R}0}:=\overline{H_{\mathrm{R}00}}$. 
Then, similarly to proof of Ref.~\cite[Example 2 in \S X.1]{rs2}, 
all self-adjoint extensions of $H_{\mathrm{L}0}$ and 
$H_{\mathrm{R}0}$ are represented with real parameters 
$\alpha_{{}_{\mathrm{L}}}$ and $\alpha_{{}_{\mathrm{R}}}$ 
in the following: 
For every $\alpha_{{}_{\mathrm{L}}} \in \mathbb{R}$, 
we have the self-adjoint extension 
\begin{equation} 
H_{\alpha_{{}_{\mathrm{L}}}}= -\, \frac{d^{2}}{dx^{2}}\,\,\, 
\textrm{with}\,\,\, 
D(H_{\alpha_{{}_{\mathrm{L}}}})= 
\left\{\psi\in AC^{2}(\overline{\Omega_{\mathrm{L}}})\, |\, 
\psi^{\prime}(-\Lambda)=\alpha_{{}_{\mathrm{L}}}\psi(-\Lambda)
\right\}, 
\label{eq:H_L-1}
\end{equation}
and for $\alpha_{{}_{\mathrm{L}}}=\infty$, 
we have the self-adjoint extension 
\begin{equation} 
H_{\infty}= -\, \frac{d^{2}}{dx^{2}}\,\,\, 
\textrm{with}\,\,\, 
D(H_{\infty})= 
\left\{\psi\in AC^{2}(\overline{\Omega_{\mathrm{L}}})\, |\, 
\psi^{\prime}(-\Lambda)=0
\right\}.  
\label{eq:H_L-2}
\end{equation}
Similarly, for every $\alpha_{{}_{\mathrm{R}}} \in \mathbb{R}$, 
we have the self-adjoint extension 
\begin{equation} 
H_{\alpha_{{}_{\mathrm{R}}}}= -\, \frac{d^{2}}{dx^{2}}\,\,\, 
\textrm{with}\,\,\, 
D(H_{\alpha_{{}_{\mathrm{R}}}})= 
\left\{\psi\in AC^{2}(\overline{\Omega_{\mathrm{R}}})\, |\, 
\psi^{\prime}(\Lambda)=\alpha_{{}_{\mathrm{R}}}\psi(\Lambda)
\right\}, 
\label{eq:H_R-1}
\end{equation}
and for $\alpha_{{}_{\mathrm{R}}}=\infty$, 
we have the self-adjoint extension 
\begin{equation} 
H_{\infty}= -\, \frac{d^{2}}{dx^{2}}\,\,\, 
\textrm{with}\,\,\, 
D(H_{\infty})= 
\left\{\psi\in AC^{2}(\overline{\Omega_{\mathrm{R}}})\, |\, 
\psi^{\prime}(\Lambda)=0
\right\}.  
\label{eq:H_R-2}
\end{equation}
Here $\overline{\Omega}$ denotes the closure 
of a set $\Omega \subset \mathbb{R}$. 
\subsection{Mathematical Setups for Our Model}  
In this paper, the closed interval $[-\Lambda , \Lambda]$ 
represents a junction on the line 
for an arbitrarily fixed constant $\Lambda>0$. 
We define a one-dimensional, the non-Euclidean space $\Omega_{\Lambda}$ 
by eliminating the junction from the line 
$(-\infty , \infty)$, i.e., 
$\Omega_{\Lambda}
:=(-\infty , -\Lambda)\cup 
(\Lambda , \infty)$. 
We assume that a free Schr\"{o}dinger particle 
such as a non-relativistic electron lives in 
$\Omega_{\Lambda}$. 
To consider self-adjoint extensions $H$ as 
Hamiltonians of the particle, 
we begin with giving the action $H_{00}$ 
of the energy operator with a small 
domain $D(H_{00})$ in which 
$H_{00}$ is not self-adjoint yet since it is 
smaller than $D(H)$. 
In the next section we will show how a self-adjoint 
extension is determined so that the extension corresponds 
to the boundary condition of each physical set-up. 

We consider the Hilbert space $L^{2}(\Omega_{\Lambda})$ 
defined as the set of all square integrable functions 
on $\Omega_{\Lambda}$. 
This represents the state space to which wave functions of our 
Schr\"{o}dinger particle belong. 
The energy operator $H_{00}$ is defined by 
\begin{equation} 
H_{00}:= -\, \frac{d^{2}}{dx^{2}}\,\,\, 
\textrm{with}\,\,\, 
D(H_{00}):= C_{0}^{\infty}(\Omega_{\Lambda}).  
\label{eq:action}
\end{equation}
Then, although the operator $H_{00}$ is neither closed nor 
self-adjoint, $H_{00}$ is symmetric.  
We denote the closure of $H_{00}$ by $H_{0}$, i.e., 
$H_{0}:=\overline{H_{00}}$. 
Then, by the well-known theorem that 
$H_{0}^{*}=H_{00}^{*}$, and moreover, 
$H_{0}\subset H_{0}^{*}$. 
So, $H_{0}$ is symmetric, 
though $H_{0}^{*}$ is \textit{not} symmetric. 
Thus, $H_{0}^{*}$ has some purely imaginary eigenvalues. 
Then, as in Definition in Ref.~\cite[\S X.1]{rs2}, 
we define vector spaces $\mathcal{H}_{+}(H_{0})$ 
and $\mathcal{H}_{-}(H_{0})$ by 
$\mathcal{H}_{+}(H_{0}):=\left\{\psi\in D(H_{0}^{*})\, 
|\, H_{0}^{*}\psi=i\psi\right\}$ 
and 
$\mathcal{H}_{-}(H_{0}):=\left\{\psi\in D(H_{0}^{*})\, 
|\, H_{0}^{*}\psi=\, -i\psi\right\}$
respectively. 
We call $\mathcal{H}_{+}(H_{0})$ and 
$\mathcal{H}_{-}(H_{0})$ 
the \textit{deficiency subspaces}. 

We can respectively prove the first part of 
the following proposition in the same way as 
the proof of Ref.~\cite[Theorem 8.25(b)]{wei} 
and the second part similarly to the proof of 
Ref.~\cite[Theorem 8.22]{wei} 
(see also Ref.~\cite[Example 3 in \S VIII.6]{rs1}):  
\begin{proposition}
\label{prop:1}
The operators $H_{0}$ and $H_{0}^{*}$ have 
the following actions with the domains respectively: 
\begin{eqnarray}
& H_{0}= -\, \frac{d^{2}}{dx^{2}} \nonumber \\ 
& \textrm{with} \,\,\, 
D(H_{0})= \left\{\psi\in D(H_{0}^{*})\, |\, 
\psi(-\Lambda)=\psi(\Lambda)=\psi'(-\Lambda)=\psi'(\Lambda)=0
\right\},  
\label{eq:H_0}
\end{eqnarray} 
and 
\begin{equation}
H_{0}^{*}= -\, \frac{d^{2}}{dx^{2}}\,\,\, 
\textrm{with}\,\,\,  
D(H_{0}^{*})= \left\{\psi\in L^{2}(\Omega_{\Lambda})\, |\, 
\psi \in AC^{2}(\overline{\Omega_{\Lambda}})
\right\}.   
\label{eq:H_0^*}
\end{equation} 
\end{proposition}

Theorem X.2 of Ref.~\cite{rs2}, together with its corollary and 
Proposition \ref{prop:1}, says that 
for every self-adjoint extension $H_{{}_{U}}$ of $H_{0}$ 
there is a unitary operator $U:\mathcal{H}_{+}(H_{0})
\longrightarrow \mathcal{H}_{-}(H_{0})$ so that 
$H_{{}_{U}}=-d^{2}/dx^{2}$ with the domain:
\begin{equation}
D(H_{{}_{U}})=
\left\{\psi_{0}+\psi_{+}+U\psi_{+}\, |\, 
\psi_{0}\in D(H_{0}), \psi_{+}\in\mathcal{H}_{+}(H_{0})
\right\}.
\label{eq:sae-domain}
\end{equation}
Conversely, for every unitary operator 
$U:\mathcal{H}_{+}(H_{0})
\longrightarrow \mathcal{H}_{-}(H_{0})$ 
the operator $H_{{}_{U}}=- d^{2}/dx^{2}$ with 
the domain given by Eq. (\ref{eq:sae-domain}) 
is a self-adjoint extension of $H_{0}$. 
That is, the self-adjoint extensions $H_{{}_{U}}$ 
of $H_{0}$ are in one-to-one correspondence 
with the set of all unitary operators 
$U:\mathcal{H}_{+}(H_{0})
\longrightarrow \mathcal{H}_{-}(H_{0})$. 

Solving simple differential equations, 
we can obtain the eigenfunctions $R_{\pm}$ and $L_{\pm}$ of 
$H_{0}^{*}$:
\begin{eqnarray}
&
R_{+}(x):= \left\{ 
\begin{array}{cl}
0 & \textrm{if} \ -\infty< x < \Lambda, \\ 
Ne^{(-1+i)x/\sqrt{2}} & \textrm{if} \ \Lambda<x<\infty, \\ 
\end{array} \right. \\ 
&
R_{-}(x):= \left\{ 
\begin{array}{cl}
0 & \textrm{if} \ -\infty<x<\Lambda, \\ 
Ne^{(-1-i)x/\sqrt{2}} & \textrm{if} \ \Lambda<x<\infty, \\ 
\end{array} \right.
\end{eqnarray}
and 
\begin{eqnarray}
&
L_{+}(x):= \left\{ 
\begin{array}{cl}
Ne^{(1-i)x/\sqrt{2}} & \textrm{if} -\infty<x<\Lambda, \\ 
0 & \textrm{if} \ \Lambda<x<\infty, \\ 
\end{array} \right. \\ 
&
L_{-}(x):= \left\{ 
\begin{array}{cl}
Ne^{(1+i)x/\sqrt{2}} & \textrm{if} \ -\infty<x<\Lambda, \\ 
0 & \textrm{if} \ \Lambda<x<\infty, \\ 
\end{array} \right.  
\end{eqnarray}
with the normalization factor $N={}^{4}\!\!\!\sqrt{2}e^{\Lambda/\sqrt{2}}$ 
so that  
$H_{0}^{*}R_{\pm}=\pm iR_{\pm}$ 
and  
$H_{0}^{*}L_{\pm}=\pm iL_{\pm}$. 
Namely, $L_{+}, R_{+}\in\mathcal{H}_{+}(H_{0})$ and 
$L_{-}, R_{-}\in\mathcal{H}_{-}(H_{0})$. 
The uniqueness of the differential equations tells us that 
\begin{eqnarray}
&
\mathcal{H}_{+}(H_{0})=
\left\{
c_{{}_{\mathrm{L}}}L_{+}+c_{{}_{\mathrm{R}}}R_{+}\, |\, 
c_{{}_{\mathrm{L}}}, c_{{}_{\mathrm{R}}}\in\mathbb{C}
\right\}, 
\\ 
&
\mathcal{H}_{-}(H_{0})=
\left\{
c_{{}_{\mathrm{L}}}L_{-}+c_{{}_{\mathrm{R}}}R_{-}\, |\, 
c_{{}_{\mathrm{L}}}, c_{{}_{\mathrm{R}}}\in\mathbb{C}
\right\},  
\end{eqnarray}
and thus, the dimensions of $\mathcal{H}_{+}(H_{0})$ 
and $\mathcal{H}_{-}(H_{0})$ are given as 
$\mathrm{dim}\, \mathcal{H}_{+}(H_{0}) = 2 
= \mathrm{dim}\, \mathcal{H}_{-}(H_{0})$, respectively~\footnote{The densely 
defined symmetric operators can be classified by the 
{\it deficiency theorem} ( see Refs.~\cite{BFV} and ~\cite[Appendix B]{SH} for physicists) 
using the dimensions of the deficiency subspaces. In the case of 
$\mathrm{dim}\, \mathcal{H}_{+}(H_{0}) = \mathrm{dim}\, \mathcal{H}_{-}(H_{0})$, 
$H_0$ has a self-adjoint extension due to the deficiency theorem.}. 
This says that the set of all unitary operators 
$U:\mathcal{H}_{+}(H_{0})
\longrightarrow\mathcal{H}_{-}(H_{0})$ makes $SU(2)$, 
and thus, that the unitary operator $U: \mathcal{H}_{+}(H_{0})
\longrightarrow \mathcal{H}_{-}(H_{0})$ is given either by 
\begin{equation}
UL_{+}=\gamma_{{}_{\mathrm{L}}}L_{-}\,\,\, 
\textrm{and}\,\,\, 
UR_{+}=\gamma_{{}_{\mathrm{L}}}R_{-}
\label{eq:diagonal}
\end{equation}
for some $\gamma_{{}_{\mathrm{L}}}, \gamma_{{}_{\mathrm{R}}} 
\in\mathbb{C}$ with $|\gamma_{{}_{\mathrm{L}}}|=1=
|\gamma_{{}_{\mathrm{R}}}|$ 
or by 
\begin{equation}
UL_{+}=\gamma_{{}_{\rightarrow}}R_{-}\,\,\, 
\textrm{and}\,\,\, 
UR_{+}=\gamma_{{}_{\leftarrow}}L_{-}
\label{eq:off-diagonal}
\end{equation}
for some $\gamma_{{}_{\rightarrow}}, \gamma_{{}_{\leftarrow}} 
\in\mathbb{C}$ with $|\gamma_{{}_{\rightarrow}}|=1=
|\gamma_{{}_{\leftarrow}}|$. 
Let us denote the vector 
$(\gamma_{{}_{\mathrm{L}}} , \gamma_{{}_{\mathrm{R}}})$ 
or $(\gamma_{{}_{\rightarrow}} , \gamma_{{}_{\leftarrow}})$ by $\gamma$. 
Then, using the one-to-one correspondence 
$U\longleftrightarrow\gamma$ given by 
Eqs. (\ref{eq:diagonal}) and (\ref{eq:off-diagonal}), 
we can represent $H_{{}_{U}}$ by $H_{\gamma}$. 
Thus, seeking a self-adjoint extension of $H_{0}$ 
is equivalent to finding a $H_{\gamma}$ with $H_{0}\subset 
H_{\gamma}=H_{\gamma}^{*}\subset H_{0}^{*}$ 
for a unitary operator $U$, that is, a vector 
$\gamma=(\gamma_{{}_{\mathrm{L}}} , \gamma_{{}_{\mathrm{R}}})$ 
or $\gamma=(\gamma_{{}_{\rightarrow}} , \gamma_{{}_{\leftarrow}})$. 

To find a boundary condition that a self-adjoint extension 
$H_{\gamma}$ of $H_{0}$ satisfies, 
we use the following tool: 
\begin{equation}
W(\varphi , \phi):= W_{-\Lambda}(\varphi^{*} , \phi) 
- W_{\Lambda}(\varphi^{*} , \phi)  
\label{eq:W}
\end{equation}
for all vectors $\varphi, \phi \in D(H_{0}^{*})$, 
where $W_{x}(f,g)$ is the Wronskian: 
$W_{x}(f,g):=f^{\prime}(x)g(x)-f(x)g^{\prime}(x)$.  
\section{Phase Factor in Boundary Conditions} 
\label{section:pf} 
In this section we investigate boundary conditions 
when the Schr\"{o}dinger particle both does and does not 
tunnel through the junction. 

\subsection{Non-Tunneling Schr\"{o}dinger Particle}
\label{subsec:non-tunneling} 
Following Eqs.(\ref{eq:sae-domain}) and (\ref{eq:diagonal}), 
wave functions $\psi$ of a self-adjoint 
extension of $H_{0}$ is given as 
$\psi=\psi_{0}+(c_{{}_{\mathrm{L}}}L_{-}+c_{{}_{\mathrm{R}}}R_{-})
+(c_{{}_{\mathrm{L}}}UL_{-}+c_{{}_{\mathrm{R}}}UR_{-})$, 
where $\psi_{0}\in D(H_{0})$, 
$c_{{}_{\mathrm{L}}}, c_{{}_{\mathrm{R}}}\in\mathbb{C}$. 
Thus, in the case where $\psi$ does not tunnel through 
the junction, the unitary operator 
$U:\mathcal{H}_{+}(H_{0})\longrightarrow 
\mathcal{H}_{-}(H_{0})$ should be given by 
$UL_{+}=\gamma_{{}_{\mathrm{L}}}L_{-}$ and 
$UR_{+}=\gamma_{{}_{\mathrm{R}}}R_{-}$ 
for some $\gamma_{{}_{\mathrm{L}}}, \gamma_{{}_{\mathrm{R}}}
\in\mathbb{C}$ with $|\gamma_{{}_{\mathrm{L}}}|=1=|\gamma_{{}_{\mathrm{R}}}|$. 
Namely, wave functions $\psi$ have the form of 
\begin{equation}
\psi=\psi_{0}+c_{{}_{\mathrm{L}}}(L_{+}+\gamma_{{}_{\mathrm{L}}}L_{-})
+c_{{}_{\mathrm{R}}}(R_{+}+\gamma_{{}_{\mathrm{R}}}R_{-}), 
\label{eq:non-tunneling}
\end{equation}
and moreover, 
the boundary conditions of $\psi(-\Lambda)$ 
and $\psi^{\prime}(-\Lambda)$ are independent of 
those of $\psi(\Lambda)$ and $\psi'(\Lambda)$.    
Because any wave function $\psi_{{}_{\mathrm{L}}}$ on the island 
$(-\infty,-\Lambda)$ and 
any wave function $\psi_{{}_{\mathrm{R}}}$ on the island 
$(\Lambda,\infty)$ are isolated from each other. 
In this case, $\psi$ has to be mathematically equivalent 
to $\psi \cong \psi_{{}_{\mathrm{L}}}\oplus\psi_{{}_{\mathrm{R}}}$
with $\psi_{{}_{\mathrm{L}}}=\psi_{{}_{\mathrm{L}0}}
+c_{{}_{\mathrm{L}}}(L_{+}+\gamma_{{}_{\mathrm{L}}}L_{-})$ 
and 
$\psi_{{}_{\mathrm{R}}}=\psi_{{}_{\mathrm{R}0}}
+c_{{}_{\mathrm{R}}}(R_{+}+\gamma_{{}_{\mathrm{R}}}R_{-})$. 
Here we note that there are wave functions 
$\psi_{{}_{\mathrm{L}0}}\in D(H_{\mathrm{L}0})$ 
and $\psi_{{}_{\mathrm{R}0}}\in D(H_{\mathrm{R}0})$ 
so that 
$\psi_{0}= \psi_{{}_{\mathrm{L}0}}\oplus\psi_{{}_{\mathrm{R}0}}$.    
Thus, any self-adjoint extension 
$H_{\alpha}$ without tunneling 
should be divided into the Schr\"{o}dinger 
operators $H_{\alpha_{{}_{\mathrm{L}}}}$ and $H_{\alpha_{{}_{\mathrm{R}}}}$ as follows: 
\begin{equation}
H_{\alpha} \cong 
H_{\alpha_{{}_{\mathrm{L}}}}\oplus H_{\alpha_{{}_{\mathrm{R}}}}, 
\end{equation}
using self-adjoint extensions shown in 
Eqs. (\ref{eq:H_L-1}), (\ref{eq:H_L-2}), (\ref{eq:H_R-1}), 
and (\ref{eq:H_R-2}), 
where $I$ denotes the identity operator.   

The following theorem and proposition establish the above physical image. 
Namely, we can classify self-adjoint extensions of 
$H_{0}$ of which wave functions cannot tunnel through the junction 
in the following: 
\begin{theorem}
\label{theo:1}
\begin{enumerate}
\item Define the action of the Hamiltonian $H_{\gamma}$ by 
$H_{\gamma}:= -d^{2}/dx^{2}$ with 
$\gamma:=(\gamma_{{}_{\mathrm{L}}} , \gamma_{{}_{\mathrm{R}}})$, 
and give its domain $D(H_{\gamma})$ by the set of all 
wave functions $\psi$ satisfying Eq. (\ref{eq:non-tunneling}):
\begin{eqnarray}
D(H_{\gamma}):=
& \left\{
\psi_{0}+c_{{}_{\mathrm{L}}}(L_{+}+\gamma_{{}_{\mathrm{L}}}L_{-})
+c_{{}_{\mathrm{R}}}(R_{+}+\gamma_{{}_{\mathrm{R}}}R_{-})\,\,\, | \right. \nonumber \\
& \left. \ \ \ \ \ \ \ \ \ \ \ \ \ \ \ \ \ \ \ \ \ \ \ \ \ \ \ \ \ \ \ \ 
\psi_{0}\in D(H_{0}), \, c_{{}_{\mathrm{L}}}, \, c_{{}_{\mathrm{R}}} \in \mathbb{C} \right\}. 
\end{eqnarray}
If $\gamma_{{}_{\mathrm{L}}}$ and $\gamma_{{}_{\mathrm{R}}}$ are given 
by $\gamma_{{}_{\mathrm{L}}}:=e^{i(\theta_{{}_{\mathrm{L}}} + \sqrt{2}\Lambda)}$ 
and 
$\gamma_{{}_{\mathrm{R}}}:=e^{i(\theta_{{}_{\mathrm{R}}} + \sqrt{2}\Lambda)}$ 
for $0\le \theta_{{}_{\mathrm{L}}}, \theta_{{}_{\mathrm{R}}} <2\pi$, 
then $H_{\gamma}$ is a self-adjoint extension of $H_{0}$. \label{theo:1:1}
\item Define the action of the Hamiltonian $H_{\alpha}$ by 
$H_{\alpha}:= -d^{2}/dx^{2}$ with 
$\alpha:=(\alpha_{{}_{\mathrm{L}}} , \alpha_{{}_{\mathrm{R}}})$, 
where $\alpha_{{}_{\mathrm{L}}}, \alpha_{{}_{\mathrm{R}}}\in\mathbb{R}
\cup\left\{\infty\right\}$. 
If the domain $D(H_{\alpha})$ is given by one of (a)--(d): 
\begin{enumerate}
\item for $\alpha\in 
\mathbb{R}\times\mathbb{R}$,  
\begin{equation}
D(H_{\alpha}):= 
\left\{
\psi\in D(H_{0}^{*})\, |\, 
\psi^{\prime}(-\Lambda)=\alpha_{{}_{\mathrm{L}}}\psi(-\Lambda)\,\,\, 
\textrm{and}\,\,\, 
\psi^{\prime}(\Lambda)=\alpha_{{}_{\mathrm{R}}}\psi(\Lambda) 
\right\};  
\end{equation} \label{theo:1:2:1}
\item for $\alpha\in 
\mathbb{R}\times\left\{\infty\right\}$, 
\begin{equation}
D(H_{\alpha}):= 
\left\{
\psi\in D(H_{0}^{*})\, |\, 
\psi^{\prime}(-\Lambda)=\alpha_{{}_{\mathrm{L}}}\psi(-\Lambda)\,\,\, 
\textrm{and}\,\,\, 
\psi^{\prime}(\Lambda)=0 
\right\};  
\end{equation} \label{theo:1:2:2}
\item for $\alpha\in 
\left\{\infty\right\}\times\mathbb{R}$, 
\begin{equation}
D(H_{\alpha}):= 
\left\{
\psi\in D(H_{0}^{*})\, |\, 
\psi^{\prime}(-\Lambda)=0\,\,\, 
\textrm{and}\,\,\, 
\psi^{\prime}(\Lambda)=\alpha_{{}_{\mathrm{R}}}\psi(\Lambda) 
\right\};  
\end{equation} \label{theo:1:2:3}
\item for $\alpha=(\infty  , \infty)$,
\begin{equation}
D(H_{\alpha}):= 
\left\{
\psi\in D(H_{0}^{*})\, |\, 
\psi^{\prime}(-\Lambda)=0=\psi^{\prime}(\Lambda) 
\right\};  
\end{equation} \label{theo:1:2:4}
\end{enumerate}
then $H_{\alpha}$ is a self-adjoint extension of $H_{0}$. \label{theo:1:2}
\item Every self-adjoint extension 
$H_{\gamma}$ as in part (\ref{theo:1:1}) and every 
self-adjoint extension $H_{\alpha}$ as in 
part (\ref{theo:1:2}) become equal to each other (i.e., $H_{\alpha}=H_{\gamma}$) 
with the one-to-one correspondence, 
$(\mathbb{C}\cup\left\{\infty\right\})\times
(\mathbb{C}\cup\left\{\infty\right\})\longrightarrow 
[0 , 2\pi)\times [0 , 2\pi)$ : 
\begin{equation}
\alpha_{{}_{\mathrm{L}}}=
\frac{1+\cos\theta_{{}_{\mathrm{L}}} - \sin\theta_{{}_{\mathrm{L}}}}{
\sqrt{2}(1+\cos\theta_{{}_{\mathrm{L}}})}\,\,\,\,\, 
\textrm{and}\,\,\,\,\, 
\alpha_{{}_{\mathrm{R}}}= 
-\, 
\frac{1+\cos\theta_{{}_{\mathrm{R}}} - \sin\theta_{{}_{\mathrm{R}}}}{
\sqrt{2}(1+\cos\theta_{{}_{\mathrm{R}}})},
\end{equation}
where $\alpha_{{}_{\sharp}}=\infty$ if 
$\theta_{{}_{\sharp}}=\pi$, $\sharp=\mathrm{L}, \,
\mathrm{R}$. Therefore, $H_{\alpha}$ is equivalent to 
$H_{\alpha_{{}_{\mathrm{L}}}}\oplus H_{\alpha_{{}_{\mathrm{R}}}}$, i.e., 
\begin{equation}
H_{\alpha} \cong 
H_{\alpha_{{}_{\mathrm{L}}}}\oplus H_{\alpha_{{}_{\mathrm{R}}}}.
\end{equation} \label{theo:1:3}
\end{enumerate}
\end{theorem}
Before proving Theorem \ref{theo:1}, 
the following lemma which is easily proved in the same way as in Ref.~\cite[Example 2 in \S X.1]{rs2}:
\begin{lemma}
\label{lem:1}
Let $\theta_{{}_{\mathrm{L}}}, 
\theta_{{}_{\mathrm{R}}}\in 
[0 , \pi)\cup(\pi , 2\pi)$ 
and $\alpha_{{}_{\mathrm{L}}}, \alpha_{{}_{\mathrm{R}}} 
\in\mathbb{R}$ be arbitrarily given. 
Then, any subspace $D(H_{\gamma})$ 
as in Theorem \ref{theo:1}~(\ref{theo:1:1}) and any subspace 
$D(H_{\alpha})$ as in Theorem \ref{theo:1}~(\ref{theo:1:2:1})
are equal if and only if 
the following correspondence holds:  
\begin{eqnarray}
\left\{ \begin{array}{l}
\sqrt{2}(1+\cos\theta_{{}_{\mathrm{L}}})\alpha_{{}_{\mathrm{L}}}
=1+\cos\theta_{{}_{\mathrm{L}}}-\sin\theta_{{}_{\mathrm{L}}}, \\ 
\sqrt{2}(1+\cos\theta_{{}_{\mathrm{R}}})\alpha_{{}_{\mathrm{R}}}
=\, -\, \left(
1+\cos\theta_{{}_{\mathrm{R}}}-\sin\theta_{{}_{\mathrm{R}}}
\right), 
\end{array} \right.
\label{eq:correspondence-isolated}
\end{eqnarray}
where
\begin{equation}
\gamma_{{}_{\mathrm{L}}}=e^{i(\theta_{{}_{\mathrm{L}}}+\sqrt{2}\Lambda)}\,\,\, 
\textrm{and}\,\,\,  
\gamma_{{}_{\mathrm{R}}}=e^{i(\theta_{{}_{\mathrm{R}}}+\sqrt{2}\Lambda)}.
\end{equation}
\end{lemma}
\begin{proof} 
Assume $D(H_{\gamma})=D(H_{\alpha})$. 
Take an arbitrary vector $\psi\in D(H_{\gamma})$. 
It is equivalent to take the vector 
$\psi=\psi_{0}+c_{{}_{\mathrm{L}}}L_{+}+c_{{}_{\mathrm{R}}}R_{+}
+c_{{}_{\mathrm{L}}}\gamma_{{}_{\mathrm{L}}}L_{-}
+c_{{}_{\mathrm{R}}}\gamma_{{}_{\mathrm{R}}}R_{-}$ 
for arbitrary $c_{{}_{\mathrm{L}}}, c_{{}_{\mathrm{R}}}\in
\mathbb{C}$ and arbitrary $\psi_{0}\in D(H_{0})$.
By the boundary condition as in Theorem \ref{theo:1}~(\ref{theo:1:2:1}), 
we have 
\begin{eqnarray}
\left\{ \begin{array}{l}
\frac{1-i}{\sqrt{2}}c_{{}_{\mathrm{L}}}L_{+}(-\Lambda)
+\frac{1+i}{\sqrt{2}}c_{{}_{\mathrm{L}}}\gamma_{{}_{\mathrm{L}}}L_{-}(-\Lambda)
= \alpha_{{}_{\mathrm{L}}}c_{{}_{\mathrm{L}}}L_{+}(-\Lambda)
+ \alpha_{{}_{\mathrm{L}}}c_{{}_{\mathrm{L}}}\gamma_{{}_{\mathrm{L}}}L_{-}(-\Lambda), \\ 
\frac{-1+i}{\sqrt{2}}c_{{}_{\mathrm{R}}}R_{+}(\Lambda)
+\frac{-1-i}{\sqrt{2}}c_{{}_{\mathrm{R}}}\gamma_{{}_{\mathrm{R}}}R_{-}(\Lambda)
= \alpha_{{}_{\mathrm{R}}}c_{{}_{\mathrm{R}}}R_{+}(\Lambda)
+ \alpha_{{}_{\mathrm{R}}}c_{{}_{\mathrm{R}}}\gamma_{{}_{\mathrm{R}}}R_{-}(\Lambda). 
\end{array} \right.
\label{eq:lem1-1}
\end{eqnarray}
It should be noted that 
\begin{eqnarray}
\left\{ \begin{array}{l}
L_{+}(-\Lambda)=R_{+}(\Lambda), \\ 
L_{-}(-\Lambda)=R_{-}(\Lambda)=R_{+}(\Lambda)^{*},
\end{array} \right.
\label{eq:LR-property}
\end{eqnarray}
and 
\begin{equation}
R_{+}(\Lambda)=R_{+}(\Lambda)^{*}e^{i\sqrt{2}\Lambda}.
\label{eq:RR^*-property}
\end{equation}
Using Eqs. (\ref{eq:lem1-1}) -- (\ref{eq:RR^*-property}), 
since $c_{{}_{\mathrm{L}}}$ and $c_{{}_{\mathrm{R}}}$ are arbitrary, 
we obtain 
\begin{equation}
\left\{ \begin{array}{l}
\frac{1-i}{\sqrt{2}}+\frac{1+i}{\sqrt{2}}e^{i\theta_{{}_{\mathrm{L}}}} 
=\alpha_{{}_{\mathrm{L}}}(1+e^{i\theta_{{}_{\mathrm{L}}}}), \\ 
\frac{-1+i}{\sqrt{2}}
+\frac{-1-i}{\sqrt{2}}e^{i\theta_{{}_{\mathrm{R}}}} 
=\alpha_{{}_{\mathrm{R}}}(1+e^{i\theta_{{}_{\mathrm{R}}}}), 
\end{array} \right.
\end{equation}
which leads to Eq.(\ref{eq:correspondence-isolated}). 

Conversely, it is easy to check that 
Eq. (\ref{eq:correspondence-isolated}) implies 
the equality $D(H_{\gamma})=D(H_{\alpha})$. 
\end{proof}
\begin{proof}[Proof of Theorem~\ref{theo:1}]
The part (\ref{theo:1:1}) just follows from 
Eqs. (\ref{eq:sae-domain}) and (\ref{eq:diagonal}). 

We employ the same method as in Ref.~\cite[Theorem 8.26]{wei} 
to prove the part (\ref{theo:1:2}). 
Note that $D(H_{0})\subsetneq D(H_{\alpha}) 
\subsetneq D(H_{0}^{*})$ and that $\langle\varphi|-\phi''\rangle
=\langle- \varphi''|\phi\rangle 
+ W(\varphi , \phi)$ 
for all $\varphi, \phi \in D(H_{0}^{*})$ first. 
Simple calculations lead to the fact that $W(\varphi , \phi)=0$ 
for all $\varphi, \phi \in D(H_{\alpha})$ given by one of (a) -- (d). 
Thus, $H_{\alpha}$ is symmetric, i.e., $H_{\alpha}\subset H_{\alpha}^{*}$. 
Let $\varphi \in D(H_{\alpha}^{*})$. 
Then, $\langle H_{\alpha}^{*}\varphi|\phi\rangle 
= \langle\varphi|H_{\alpha}\phi\rangle$ 
for every $\phi\in D(H_{\alpha})$. 
Thus, $\langle-\varphi''|\phi\rangle 
= \langle\varphi|-\phi''\rangle$ 
by Proposition~\ref{prop:1} since $H_{\alpha}\subset 
H_{\alpha}^{*}\subset H_{0}^{*}$. 
It means that $W(\varphi , \phi)=0$. 
Take any function $\phi\in D(H_{\alpha})$ 
with $\phi(-\Lambda)\ne 0$ and $\phi(\Lambda)=0$. 
Then, using $\phi$ with the boundary condition 
in the domain (a) we have $\varphi'(-\Lambda)=
\alpha_{{}_{\mathrm{L}}}\varphi(-\Lambda)$. 
Similarly, using a function $\phi\in D(H_{\alpha})$ 
with $\phi(\Lambda)\ne 0$ and $\phi(-\Lambda)=0$ 
we reach the fact that $\varphi'(\Lambda)=
\alpha_{{}_{\mathrm{R}}}\varphi(\Lambda)$. 
Thus, $\varphi\in D(H_{\alpha})$, that is, 
$H_{\alpha}^{*}\subset H_{\alpha}$. 
Therefore, $H_{\alpha}$ is self-adjoint. 
Since we can similarly handle the other cases, 
we complete the proof of the part (\ref{theo:1:2}). 

The part (\ref{theo:1:3}) for the case (\ref{theo:1:2:1}) 
directly follows from Lemma \ref{lem:1}. 
For $\alpha\in\mathbb{R}\times\left\{\infty\right\}$ in 
the case (\ref{theo:1:2:2}), we expand Eq. (\ref{eq:correspondence-isolated}) 
to the case $\alpha_{{}_{\mathrm{R}}}=\infty$ and 
$\theta_{{}_{\mathrm{R}}}=\pi$ as  
$0 \infty=0 \alpha_{{}_{\mathrm{R}}} = 0$. 
For $\alpha\in\left\{\infty\right\}\times\mathbb{R}$ in 
the case (\ref{theo:1:2:3}) and 
$\alpha=(\infty , \infty)$ in the case (\ref{theo:1:2:4}), 
we expand Eq. (\ref{eq:correspondence-isolated}) 
in the same way as in the case (\ref{theo:1:2:2}). 
These arguments complete the part (\ref{theo:1:3}).
\end{proof}
\begin{proposition}
\label{prop:2}
For every vector $\alpha=(\alpha_{{}_{\mathrm{L}}} , 
\alpha_{{}_{\mathrm{R}}})\in(\mathbb{R}\cup\left\{\infty\right\})
\times(\mathbb{R}\cup\left\{\infty\right\})$, 
there is no vector $\gamma=(\gamma_{{}_{\rightarrow}} , 
\gamma_{{}_{\leftarrow}})$ given in Eq. (\ref{eq:off-diagonal}) so that 
$D(H_{\alpha}) = D(H_{\gamma})$. 
Conversely, for every vector $\gamma=(\gamma_{{}_{\rightarrow}} , 
\gamma_{{}_{\leftarrow}})$ given in Eq. (\ref{eq:off-diagonal}), 
there is no vector 
$\alpha=(\alpha_{{}_{\mathrm{L}}} , 
\alpha_{{}_{\mathrm{R}}})\in(\mathbb{R}\cup\left\{\infty\right\})
\times(\mathbb{R}\cup\left\{\infty\right\})$ so that 
$D(H_{\alpha}) = D(H_{\gamma})$. 
\end{proposition}
\begin{proof}
Let a vector $\alpha=(\alpha_{{}_{\mathrm{L}}} , 
\alpha_{{}_{\mathrm{R}}})$ be in $(\mathbb{R}\cup\left\{\infty\right\})
\times(\mathbb{R}\cup\left\{\infty\right\})$. 
Suppose for the sake of contradiction that there is a vector $\gamma=(\gamma_{{}_{\rightarrow}} , 
\gamma_{{}_{\leftarrow}})$ given in Eq. (\ref{eq:off-diagonal}) so that 
$D(H_{\alpha}) = D(H_{\gamma})$. 
Then, simple calculations lead to a contradiction; 
$e^{-i\pi/4}=\alpha_{{}_{\mathrm{L}}}=\alpha_{{}_{\mathrm{R}}}=e^{i\pi/4}$.
Therefore, we prove the first part. In the same way, we can prove the last part. 
\end{proof}
\begin{remark}
\label{rem:non-tunneling} 
We assume that the wave function $\psi(x)$ is 
in $D(H_{0}^{*})$ for every $\Lambda >0$ so that 
\begin{equation}
\left\{ \begin{array}{l}
\lim_{\Lambda\to 0}\psi(-\Lambda)=\psi(0-)=\psi(0)=
\psi(0+)=\lim_{\Lambda\to 0}\psi(+\Lambda), \\
\lim_{\Lambda\to 0}\psi'(-\Lambda)=\psi'(0-), \\ 
\lim_{\Lambda\to 0}\psi'(+\Lambda)=\psi'(0+).
\end{array} \right.
\end{equation}
Let $\alpha$, $\alpha_{{}_{\mathrm{L}}}$, and 
$\alpha_{{}_{\mathrm{R}}}$ be given as 
$-\infty < \alpha, \alpha_{{}_{\mathrm{R}}}\le +\infty$ 
and $-\infty < \alpha_{{}_{\mathrm{L}}}< +\infty$. 
When $\alpha_{{}_{\mathrm{L}}}$ and $\alpha_{{}_{\mathrm{R}}}$ 
are arbitrarily given, we define $\alpha$ 
by $\alpha :=\alpha_{{}_{\mathrm{R}}}-\alpha_{{}_{\mathrm{L}}}$. 
Conversely, when $\alpha$ is arbitrarily given, 
we divide $\alpha$ into $\alpha_{{}_{\mathrm{L}}}$ 
and $\alpha_{{}_{\mathrm{R}}}$ 
as $\alpha =\alpha_{{}_{\mathrm{R}}}-\alpha_{{}_{\mathrm{L}}}$. 
Let us assume that the boundary condition in Theorem \ref{theo:1} 
holds for all $\Lambda>0$. 
Then, our boundary condition in Theorem \ref{theo:1} 
tends to the boundary condition in Ref.~\cite[Eq.(3.1.9)]{AGH-KHP} for the point interaction 
as $\Lambda \to 0$: 
\begin{equation}
\psi^{\prime}(0+)-\psi^{\prime}(0-)= \alpha \, \psi(0). 
\end{equation}
\end{remark}
Theorem \ref{theo:1} says that 
there is no phase factor in the boundary condition 
when every wave function does not 
tunnel through the junction. 
It should be noted that all the cases of Eq. (\ref{eq:non-tunneling}) 
are described as in part (\ref{theo:1:1}) of Theorem \ref{theo:1} 
since $\theta_{{}_{\mathrm{L}}}$ and $\theta_{{}_{\mathrm{R}}}$ 
are independent. 
On the other hand, in the case where some wave functions 
tunnel through the junction, we find another type 
of the boundary conditions, and then, 
we realize that some phase factors are in this type. 
We will show this in the next subsection.  
\subsection{Tunneling Schr\"{o}dinger Particle}
\label{subsec:tunneling} 
In this subsection, we show that 
the Schr\"{o}dinger particle 
tunneling the junction comes up with 
another type of the boundary conditions (see Definition~\ref{def:2}). 
Following Eq.~(\ref{eq:sae-domain}) again, 
the unitary operator 
$U:\mathcal{H}_{+}(H_{0})\longrightarrow 
\mathcal{H}_{-}(H_{0})$ should be given by 
\begin{equation}
UL_{+}=\gamma_{{}_{\rightarrow}}R_{-}\quad \textrm{and}\quad 
UR_{+}=\gamma_{{}_{\leftarrow}}L_{-}
\label{eq:crossing}
\end{equation} 
for some $\gamma_{{}_{\rightarrow}}, \gamma_{{}_{\leftarrow}}
\in\mathbb{C}$ with $|\gamma_{{}_{\rightarrow}}|=1
=|\gamma_{{}_{\leftarrow}}|$.  
Namely, all wave functions $\psi$ of any self-adjoint 
extension of $H_{0}$ satisfying Eq. (\ref{eq:crossing}) 
have the following form:
\begin{equation}
\psi=\psi_{0}+c_{{}_{\mathrm{L}}}(L_{+}+\gamma_{{}_{\rightarrow}}R_{-})
+c_{{}_{\mathrm{R}}}(R_{+}+\gamma_{{}_{\leftarrow}}L_{-}), 
\label{eq:tunneling}
\end{equation}
and moreover, 
the boundary conditions of $\psi(\Lambda)$ 
and $\psi^{\prime}(\Lambda)$ are dependent on 
those of $\psi(-\Lambda)$ and $\psi^{\prime}(-\Lambda)$.    
For the wave functions with the form of 
Eq.(\ref{eq:tunneling}), 
we can find a phase factor 
in some boundary conditions. 

We define some mathematical notions: 
\begin{definition}
\label{def:1}
A vector $\vec{\alpha}=
(\alpha_{1} , \alpha_{2} , \alpha_{3} , \alpha_{4}) 
\in \mathbb{C}$ belongs to the class $\mathcal{A}$ 
(i.e., $\vec{\alpha}\in\mathcal{A}$) if 
the vector $\vec{\alpha}$ satisfies 
\begin{description}
\item[($\mathcal{A}$1)] $\alpha_{1}\alpha_{4}^{*}-\alpha_{2}^{*}\alpha_{3}=1$;  
\item[($\mathcal{A}$2)] $\alpha_{1}\alpha_{3}^{*}$, $\alpha_{2}\alpha_{4}^{*} \in \mathbb{R}$.
\end{description}
\end{definition}
\begin{definition}
\label{def:2}
Fix $\vec{\alpha}=
(\alpha_{1} , \alpha_{2} , \alpha_{3} , \alpha_{4}) 
\in \mathbb{C}$. 
Then, a wave function $\psi\in D(H_{0}^{*})$ satisfies 
the boundary conditions $BC (\vec{\alpha})$ if 
the wave function $\psi$ satisfies 
\begin{equation}
\left\{ \begin{array}{l}
\psi(\Lambda)=\alpha_{1}\psi(-\Lambda)+\alpha_{2}\psi'(-\Lambda), \\ 
\psi^{\prime}(\Lambda)=\alpha_{3}\psi(-\Lambda)+\alpha_{4}\psi'(-\Lambda).
\end{array} \right.
\end{equation}
\end{definition}
We define a function $F:\mathbb{C}^{2}\longrightarrow\mathbb{C}$ 
by 
\begin{equation}
F(z_{1} , z_{2}):= 
|z_{1}|^{2}+\sqrt{2}z_{1}z_{2} +|z_{2}|^{2}-1
\end{equation}
for every $(z_{1} , z_{2}) \in\mathbb{C}^{2}$. 
\begin{definition}
\label{def:3}
A vector $\vec{\alpha}=
(\alpha_{1} , \alpha_{2} , \alpha_{3} , \alpha_{4}) 
\in \mathbb{C}$ is a solution of the system $\mathcal{S}$ 
if the vector $\vec{\alpha}$ satisfies 
\begin{description}
\item[($\mathcal{S}$1)] $F(\alpha_{1} , \alpha_{2}) = 0 
= F(\alpha_{3} , \alpha_{4})$; 
\item[($\mathcal{S}$2)] ${\displaystyle \alpha_{1}+\alpha_{2}e^{-i\pi/4} 
=\left(\alpha_{3}+\alpha_{4}e^{-i\pi/4}\right)e^{i3\pi/4}}$; 
\item[($\mathcal{S}$3)] ${\displaystyle \alpha_{1}+\alpha_{2}e^{i\pi/4} 
=\left(\alpha_{3}+\alpha_{4}e^{i\pi/4}\right)e^{-i3\pi/4}}$. 
\end{description}
\end{definition}
\begin{example}
\label{ex:1}
Fix $\theta$ with $0\le\theta<2\pi$ arbitrarily. 
Set $\vec{\alpha}$ as $\alpha_{1}=0$, $\alpha_{2}=-e^{i\theta}$, 
$\alpha_{3}=e^{i\theta}$, and $\alpha_{4}=0$. 
Then, the vector $\vec{\alpha}$ belongs to the class 
$\mathcal{A}$ and 
it is a solution of the system $\mathcal{S}$. 
\end{example}
This Example \ref{ex:1}, together with the following 
theorem, secures that 
the boundary condition $BC (\vec{\alpha})$ 
can include some phase factors: 
\begin{theorem}
\label{theo:4}
\begin{enumerate}
\item Fix a vector $\vec{\alpha}\in\mathcal{A}$ 
arbitrarily. Define the action of the Hamiltonian $H_{\vec{\alpha}}$ by 
$H_{\vec{\alpha}}:= -d^{2}/dx^{2}$ with 
\begin{equation}
D(H_{\vec{\alpha}}):= 
\left\{
\psi\in D(H_{0}^{*})\, |\, \textrm{$\psi$ satisfies the boundary 
condition BC($\vec{\alpha}$)}
\right\}. 
\end{equation}
Then, $H_{\vec{\alpha}}$ is a self-adjoint extension of $H_{0}$. 
\label{theo:4:1}
\item Assume the vector $\vec{\alpha}$ belongs to the class 
$\mathcal{A}$ and it is a solution of the system $\mathcal{S}$. 
Define the action of the Hamiltonian $H_{\gamma}$ by 
$H_{\gamma}:= -d^{2}/dx^{2}$ with 
$\gamma:=(\gamma_{{}_{\rightarrow}} , \gamma_{{}_{\leftarrow}})$, 
and give its domain $D(H_{\gamma})$ by the set of all 
wave functions $\psi$ satisfying Eq.(\ref{eq:non-tunneling}):
\begin{equation}
D(H_{\gamma}) :=
\left\{
\psi_{0}+c_{{}_{\mathrm{L}}}f+c_{{}_{\mathrm{R}}}g\,\,\, 
|\,\,\, \psi_{0}\in D(H_{0}), c_{{}_{\mathrm{L}}}, c_{{}_{\mathrm{R}}} 
\in \mathbb{C}
\right\}. 
\end{equation}
If $\gamma_{{}_{\mathrm{L}}}$ and $\gamma_{{}_{\mathrm{R}}}$ are given 
by
\begin{equation}
\gamma_{{}_{\rightarrow}}:=
\left(
\alpha_{1}+\alpha_{2}e^{-i\pi/4}
\right)
e^{i\sqrt{2}\Lambda}
=
\left(
\alpha_{3}+\alpha_{4}e^{-i\pi/4}
\right)
e^{i\left\{\sqrt{2}\Lambda+(3\pi/4)\right\}}, 
\end{equation}
and 
\begin{equation}
\gamma_{{}_{\leftarrow}}:=
\left(
\alpha_{1}+\alpha_{2}e^{i\pi/4}
\right)^{-1}
e^{i\sqrt{2}\Lambda}
=
\left(
\alpha_{3}+\alpha_{4}e^{i\pi/4}
\right)^{-1}
e^{i\left\{\sqrt{2}\Lambda+(3\pi/4)\right\}},
\end{equation}
then $H_{\gamma}$ is a self-adjoint extension of $H_{0}$. 
Moreover, $H_{\vec{\alpha}}=H_{\gamma}$. 
\label{theo:4:2}
\end{enumerate}
\end{theorem}
Before proving Theorem \ref{theo:4} 
we note the following five lemmas: 
\begin{lemma}
\label{lem:3}
If $\alpha_{1}, \alpha_{2}, \alpha_{3}, \alpha_{4} 
\in\mathbb{C}$ satisfy 
\begin{eqnarray}
& 
\alpha_{1}\alpha_{4}^{*}-\alpha_{2}^{*}\alpha_{3}=1, 
\label{eq:lem-3-1} \\  
& 
\alpha_{1}\alpha_{3}^{*},\,\,\, 
\alpha_{2}\alpha_{4}^{*} \in \mathbb{R}, 
\label{eq:lem-3-2}
\end{eqnarray}
then $\alpha_{j}\alpha_{j'}^{*} \in\mathbb{R}$ 
for each $j, j^{\prime}=1, 2, 3, 4$. 
\end{lemma}
\begin{proof}
In the case of $j= j^{\prime}$, 
the statement of our lemma is trivial. 
Thus, we suppose that $j\ne j^{\prime}$.  
Equation (\ref{eq:lem-3-1}) leads to 
\begin{equation}
\alpha_{3}^{*}=\alpha_{3}^{*}(\alpha_{1}\alpha_{4}^{*}-\alpha_{2}^{*}\alpha_{3})
=\alpha_{1}\alpha_{3}^{*}\alpha_{4}^{*}-\alpha_{2}^{*}|\alpha_{3}|^{2}.
\label{eq:lem-3-3}
\end{equation} 
Due to the condition (\ref{eq:lem-3-2}), 
multiplying $\alpha_{2}$ by both sides of this equation 
gives $\alpha_{2}\alpha_{3}^{*}\in\mathbb{R}$. 
Also this fact and Eq. (\ref{eq:lem-3-1}) 
say that $\alpha_{1}\alpha_{4}^{*} =1 + \alpha_{2}^{*}\alpha_{3} \in \mathbb{R}$ at the same time. 
Similarly, since Eq. (\ref{eq:lem-3-1}) leads to 
$\alpha_{2}=\alpha_{2}(\alpha_{1}\alpha_{4}^{*}-\alpha_{2}^{*}\alpha_{3})
=\alpha_{1}\alpha_{2}\alpha_{4}^{*}-|\alpha_{2}|^{2}\alpha_{3}$, 
we have $\alpha_{1}^{*}\alpha_{2}\in\mathbb{R}$. 
Using Eq. (\ref{eq:lem-3-3}), we reach 
$\alpha_{3}^{*}\alpha_{4}= \alpha_{1}\alpha_{3}^{*}|\alpha_{4}|^{2}
-\alpha_{2}^{*}\alpha_{4}|\alpha_{3}|^{2} \in \mathbb{R}$ by the condition (\ref{eq:lem-3-2}).
We can conclude all the results we desire from the facts that we showed above.
\end{proof}
\begin{lemma}
\label{lem:4}
If $\alpha_{1}\alpha_{2}^{*}, \, \alpha_{3}\alpha_{4}^{*}\in\mathbb{R}$, 
then 
\begin{equation}
\alpha_{1}+\alpha_{2}e^{i\pi/4}\ne 0 \quad 
\textrm{and} \quad 
\alpha_{3}+\alpha_{4}e^{i\pi/4} \ne 0.
\label{eq:lem-4-1}
\end{equation}
\end{lemma}
\begin{proof}
Suppose for the sake of contradiction that 
(i) $\alpha_{1}= - \alpha_{2}e^{i\pi/4}$ 
or 
(ii) $\alpha_{3}= -  \alpha_{4}e^{i\pi/4}$ holds. 
In the case where (i) holds, 
$\mathbb{R}\ni \alpha_{1}\alpha_{2}^{*} 
= - |\alpha_{2}|^{2}e^{i\pi/4}$, 
which is a contradiction. 
In the same way as we did now, 
we have a contradiction 
in the case where (ii) holds. 
\end{proof}
Straightforward calculations lead to the following 
two lemmas.
\begin{lemma}
\label{lem:5}
If $z_{1}, z_{2}\in\mathbb{C}$ 
satisfy $z_{1}z_{2}^{*}
=\left\{1-(|z_{1}|^{2}+|z_{2}|^{2})\right\}/\sqrt{2}$, 
then 
\begin{equation}
\Bigl|z_{1}+z_{2}e^{\pm i\pi/4}\Bigr|=1.  
\end{equation}
\end{lemma}
\begin{proof}
We have 
\begin{eqnarray}
\left( z_{1}+z_{2}e^{\pm i\pi/4} \right)
\left( z_{1}^{*}+z_{2}^{*}e^{\mp i\pi/4} \right)
& =
|z_{1}|^{2}+|z_{2}|^{2}+z_{1}z_{2}^{*}
\left( \frac{1+i}{\sqrt{2}}+\frac{1-i}{\sqrt{2}}\right) \nonumber \\ 
&=|z_{1}|^{2}+|z_{2}|^{2}+z_{1}z_{2}^{*}\sqrt{2}
=1,
\end{eqnarray}
noting $z_{1}z_{2}^{*}\in\mathbb{R}$ 
and using the assumption.
\end{proof}
\begin{lemma}
\label{lem:6}
Let $\alpha_{1}, \alpha_{2}, \alpha_{3}, \alpha_{4} 
\in\mathbb{C}$ be given so that 
Eq. (\ref{eq:lem-3-1}) and the condition (\ref{eq:lem-3-2}) hold. 
If $\varphi, \phi \in D(H_{0}^{*})$ satisfy 
the boundary condition $BC( \vec{\alpha})$, then $W(\varphi,\phi)=0$. 
\end{lemma}
\begin{proof}
It follows directly from Eq. (\ref{eq:lem-3-1}) 
and the condition (\ref{eq:lem-3-2}) that 
\begin{eqnarray}
& W(\varphi,\phi) = \nonumber \\
& \ \ \ \varphi(-\Lambda)^{*}\phi^{\prime}(-\Lambda)
\left(-1+\alpha_{1}^{*}\alpha_{4}-\alpha_{2}\alpha_{3}^{*}\right)
+ \varphi(-\Lambda)^{*}\phi(-\Lambda)
\left(\alpha_{1}^{*}\alpha_{3}-\alpha_{1}\alpha_{3}^{*}\right) \nonumber \\ 
& \ \ \ +
\varphi^{\prime}(-\Lambda)^{*}\phi(-\Lambda)
\left(\alpha_{2}^{*}\alpha_{3}+1-\alpha_{1}\alpha_{4}^{*}\right)  
+ \varphi^{\prime}(-\Lambda)^{*}\phi^{\prime}(-\Lambda)
\left(\alpha_{2}^{*}\alpha_{4}-\alpha_{2}\alpha_{4}^{*}\right) \nonumber \\
& \ \ \ = 0, 
\end{eqnarray}
since $\varphi, \phi \in D(H_{0}^{*})$ satisfy 
the boundary condition $BC (\vec{\alpha})$.
\end{proof}
We state the last of five lemmas: 
\begin{lemma}
\label{lem:7}
Let $\alpha_{1}, \alpha_{2}, \alpha_{3}, \alpha_{4} 
\in\mathbb{C}$ be given so that 
Eq.(\ref{eq:lem-4-1}) holds. 
Then, the boundary condition BC($\vec{\alpha}$) 
is equivalent to the following conditions: 
\begin{equation}
\left\{ \begin{array}{l}
\gamma_{{}_{\rightarrow}}
= \left( \alpha_{1}+\alpha_{2}e^{-i\pi/4} \right) e^{i\sqrt{2}\Lambda} 
= \left( \alpha_{3}+\alpha_{4}e^{-i\pi/4} \right) e^{i3\pi/4}e^{i\sqrt{2} \Lambda}, \\ 
\gamma_{{}_{\leftarrow}} =
\left( \alpha_{1}+\alpha_{2}e^{i\pi/4} \right)^{-1} e^{i\sqrt{2}\Lambda} 
= \left( \alpha_{3}+\alpha_{4}e^{i\pi/4} \right)^{-1} e^{i3\pi/4}e^{i\sqrt{2}\Lambda}.
\end{array} \right.
\label{eq:lem-7-1}
\end{equation} 
\end{lemma} 
\begin{proof}
Let $\psi\in D(H_{0}^{*})$ be an arbitrary 
wave function satisfying the boundary condition 
$BC (\vec{\alpha} )$. 
Then, we note that taking this $\psi$ 
is equivalent to giving $\psi$ with 
arbitrary $c_{{}_{\mathrm{L}}}, c_{{}_{\mathrm{R}}} \in \mathbb{C}$ 
so that $\psi= \psi_{0}+c_{{}_{\mathrm{L}}}(L_{+}+\gamma_{{}_{\rightarrow}}R_{-})
+c_{{}_{\mathrm{R}}}(R_{+}+\gamma_{{}_{\leftarrow}}L_{-})$ 
by Eqs. (\ref{eq:crossing}) and (\ref{eq:tunneling}). 
Thus, while the wave function $\psi$ is written at 
$x=\Lambda$ as $\psi(\Lambda)=
c_{{}_{\mathrm{L}}}\gamma_{{}_{\rightarrow}}R_{+}(\Lambda)^{*}
+c_{{}_{\mathrm{R}}}R_{+}(\Lambda)$ 
by Eq. (\ref{eq:LR-property}), we have 
\begin{eqnarray}
\psi(\Lambda)
&=\alpha_{1}\psi(-\Lambda)
+\alpha_{2}\psi^{\prime \prime}(-\Lambda) \nonumber \\ 
& = c_{{}_{\mathrm{L}}}
\left( \alpha_{1}+\alpha_{2}e^{-i\pi/4} \right)
R_{+}(\Lambda) + c_{{}_{\mathrm{R}}}
\left( \alpha_{1}+\alpha_{2}e^{i\pi/4} \right)
\gamma_{{}_{\leftarrow}}
R_{+}(\Lambda)^{*}.
\end{eqnarray}
Since $c_{{}_{\mathrm{L}}}, c_{{}_{\mathrm{R}}}$ are arbitrary, 
we obtain the first equality of 
$\gamma_{{}_{\rightarrow}}$ and $\gamma_{{}_{\leftarrow}}$ 
in Eq. (\ref{eq:lem-7-1}) individually. 
Employing the same argument 
we can express $\gamma_{{}_{\rightarrow}}$ and 
$\gamma_{{}_{\leftarrow}}$ by $\alpha_{3}, \alpha_{4}$ 
as in the second equality of Eq. (\ref{eq:lem-7-1}) individually.   

We can show directly that Eq. (\ref{eq:lem-7-1}) lead to 
the boundary condition $BC (\vec{\alpha})$ 
with some straightforward calculations.
\end{proof}
\begin{proof}[Proof of Theorem \ref{theo:4}] 
We employ the same method as in Ref.~\cite[Theorem 8.26]{wei} 
to prove the part (\ref{theo:4:1}). 
Note that, in the same way as noted in the proof of Theorem \ref{theo:1}, 
$D(H_{0})\subsetneq D(H_{\vec{\alpha}}) \subsetneq D(H_{0}^{*})$, 
and moreover, 
$\langle \varphi|-\phi^{\prime \prime}\rangle
=\langle - \varphi^{\prime \prime}| \phi \rangle 
+ W(\varphi , \phi)$ 
for all $\varphi, \phi \in D(H_{0}^{*})$. 
Let wave functions $\varphi$ and $\phi$ 
be in $D(H_{\vec{\alpha}})$. 
Then, there are $\varphi_{0}, \phi_{0} \in D(H_{0})$ and $a_{{}_{\mathrm{L}}}, a_{{}_{\mathrm{R}}}, 
b_{{}_{\mathrm{L}}}, b_{{}_{\mathrm{R}}} \in\mathbb{C}$ so that 
the wave functions $\varphi$ and $\phi$ are written 
as $\varphi=\varphi_{0}+a_{{}_{\mathrm{L}}}f+a_{{}_{\mathrm{R}}}g$ 
and $\phi=\phi_{0}+b_{{}_{\mathrm{L}}}f+b_{{}_{\mathrm{R}}}g$, respectively. 
Straightforward calculations lead to 
\begin{eqnarray}
& W (\varphi,\phi)= \nonumber \\
& a_{{}_{\mathrm{L}}}^{*}b_{{}_{\mathrm{L}}}
\left\{
-f(-\Lambda)^{*}f^{\prime}(-\Lambda)+f(\Lambda)^{*}f^{\prime}(\Lambda)
+f^{\prime}(-\Lambda)^{*}f(-\Lambda)-f'(\Lambda)^{*}f(\Lambda)
\right\} \nonumber \\ 
& +
a_{{}_{\mathrm{L}}}^{*}b_{{}_{\mathrm{R}}}
\left\{
-f(-\Lambda)^{*}g^{\prime}(-\Lambda)+f(\Lambda)^{*}g^{\prime}(\Lambda)
+f^{\prime}(-\Lambda)^{*}g(-\Lambda)-f'(\Lambda)^{*}g(\Lambda)
\right\} \nonumber \\ 
& +
a_{{}_{\mathrm{R}}}^{*}b_{{}_{\mathrm{L}}}
\left\{
-g(-\Lambda)^{*}f^{\prime}(-\Lambda)+g(\Lambda)^{*}f^{\prime}(\Lambda)
+g'(-\Lambda)^{*}f(-\Lambda)-g'(\Lambda)^{*}f(\Lambda)
\right\} \nonumber \\ 
& + a_{{}_{\mathrm{R}}}^{*}b_{{}_{\mathrm{R}}}
\left\{ -g(-\Lambda)^{*}g^{\prime}(-\Lambda)+g(\Lambda)^{*}g^{\prime}(\Lambda)
+g^{\prime}(-\Lambda)^{*}g(-\Lambda)-g^{\prime}(\Lambda)^{*}g(\Lambda) \right\}.
\end{eqnarray}
Using Eq. (\ref{eq:LR-property}), we have
\begin{eqnarray}
& \left\{ \begin{array}{ll}
f(-\Lambda)=R_{+}(\Lambda), 
& f(\Lambda)=\gamma_{{}_{\rightarrow}}R_{+}(\Lambda)^{*}, \\ 
f^{\prime}(-\Lambda)=e^{-i\pi/4}R_{+}(\Lambda), 
& f^{\prime}(\Lambda)=\gamma_{{}_{\rightarrow}}e^{-i3\pi/4}R_{+}(\Lambda)^{*}, 
\end{array} \right. \\
& \left\{ \begin{array}{ll}
g(-\Lambda)=\gamma_{{}_{\leftarrow}}R_{+}(\Lambda)^{*}, 
& g(\Lambda)=R_{+}(\Lambda), \\ 
g^{\prime}(-\Lambda)=\gamma_{{}_{\leftarrow}}e^{i\pi/4}R_{+}(\Lambda)^{*}, 
& g^{\prime}(\Lambda)=e^{i3\pi/4}R_{+}(\Lambda). 
\end{array} \right. 
\end{eqnarray} 
Inserting these values into $W(\varphi,\phi)$ obtained above, 
we have 
\begin{eqnarray}
W (\varphi, \phi)= &
\frac{a_{{}_{\mathrm{L}}}^{*}b_{{}_{\mathrm{L}}}
R_{+}(\Lambda)^{*}R_{+}(\Lambda)}{\sqrt{2}}
\left\{\right.
\left(-1+i\right)
+|\gamma_{{}_{\rightarrow}}|^{2}\left(-1-i\right) \nonumber \\ 
& \qquad\qquad\qquad\qquad\qquad\qquad 
 + \left(1+i\right)
 + |\gamma_{{}_{\rightarrow}}|^{2}\left(1-i\right)
\left.\right\} \nonumber \\ 
& + \frac{a_{{}_{\mathrm{L}}}^{*}b_{{}_{\mathrm{R}}}}{\sqrt{2}}
\left\{ \right.
R_{+}(\Lambda)^{*2}\gamma_{{}_{\leftarrow}}\left(-1-i\right)
+R_{+}(\Lambda)^{2}\gamma_{{}_{\rightarrow}}^{*}\left(-1+i\right) \nonumber \\ 
& \qquad\qquad\qquad 
+R_{+}(\Lambda)^{*2}\gamma_{{}_{\leftarrow}}\left(1+i\right)
+R_{+}(\Lambda)^{2}\gamma_{{}_{\rightarrow}}^{*}\left(1-i\right)
\left.\right\} \nonumber \\ 
& + \frac{a_{{}_{\mathrm{R}}}^{*}b_{{}_{\mathrm{L}}}}{\sqrt{2}}
\left\{\right.
R_{+}(\Lambda)^{2}\gamma_{{}_{\leftarrow}}^{*}\left(-1+i\right)
+R_{+}(\Lambda)^{*2}\gamma_{{}_{\rightarrow}}\left(-1-i\right) \nonumber \\ 
& \qquad\qquad\qquad 
+R_{+}(\Lambda)^{2}\gamma_{{}_{\leftarrow}}^{*}\left(1-i\right)
+R_{+}(\Lambda)^{*2}\gamma_{{}_{\rightarrow}}\left(1+i\right)
\left.\right\} \nonumber \\ 
& + \frac{a_{{}_{\mathrm{R}}}^{*}b_{{}_{\mathrm{R}}}
R_{+}(\Lambda)R_{+}(\Lambda)^{*}}{\sqrt{2}} \left\{\right.
|\gamma_{{}_{\rightarrow}}|^{2}\left(-1-i\right)
+\left(-1+i\right) \nonumber \\ 
& \qquad\qquad\qquad\qquad\qquad\qquad 
+|\gamma_{{}_{\rightarrow}}|^{2}\left(1-i\right)
+\left(1+i\right)
\left.\right\} \nonumber \\ 
& = 0.
\end{eqnarray}
Thus, $H_{\vec{\alpha}}$ is symmetric, i.e., 
$H_{\vec{\alpha}}\subset H_{\vec{\alpha}}^{*}$. 
Let $\varphi \in D(H_{\vec{\alpha}}^{*})$. 
Then, $\langle H_{\vec{\alpha}}^{*}\varphi|\phi\rangle 
= \langle\varphi|H_{\vec{\alpha}}\phi\rangle$ 
for every $\phi\in D(H_{\vec{\alpha}})$. 
Thus, $\langle-\varphi^{\prime \prime} | \phi\rangle 
= \langle\varphi|-\phi^{\prime \prime} \rangle$ 
by Proposition \ref{prop:1} since $H_{\vec{\alpha}}\subset 
H_{\vec{\alpha}}^{*}\subset H_{0}^{*}$. 
It means that $W(\varphi , \phi)=0$. 
Take any function $\phi\in D(H_{\alpha})$ 
with $\phi(-\Lambda)\ne 0$ and $\phi^{\prime}(-\Lambda)=0$. 
Then, using this $\phi$ with the boundary condition $BC ( \vec{\alpha} )$, 
we have 
\begin{equation}
\alpha_{3}\varphi(\Lambda)^{*}-\alpha_{1}\varphi^{\prime}(\Lambda)
=\, - \, \varphi^{\prime}(-\Lambda)^{*}.
\label{eq:theo-4-1a}
\end{equation}
In the same way, take any function $\phi\in D(H_{\alpha})$ 
with $\phi(-\Lambda)=0$ and $\phi^{\prime}(-\Lambda)\ne 0$. 
Then, using this function $\phi$ with the boundary condition $BC (\vec{\alpha})$, 
we have 
\begin{equation}
\alpha_{4}\varphi(\Lambda)^{*}-\alpha_{2}\varphi^{\prime}(\Lambda)
=\varphi(-\Lambda)^{*}.
\label{eq:theo-4-1b}
\end{equation}
It follows from Eqs. (\ref{eq:theo-4-1a}) and (\ref{eq:theo-4-1b}) 
that 
\begin{equation}
\left\{ \begin{array}{l}
-\alpha_{2}^{*}\alpha_{3}\varphi(\Lambda)^{*}
+\alpha_{1}\alpha_{2}^{*}\varphi^{\prime}(\Lambda)
=\alpha_{2}^{*}\varphi'(-\Lambda)^{*}, \\ 
\alpha_{1}^{*}\alpha_{4}\varphi(\Lambda)^{*}
-\alpha_{1}^{*}\alpha_{2}\varphi^{\prime}(\Lambda)
=\alpha_{1}^{*}\varphi(-\Lambda)^{*}. 
\end{array} \right.
\end{equation}
Summing these two equations gives us the equation:
\begin{eqnarray}
& \left(
\alpha_{1}^{*}\alpha_{4}-\alpha_{2}^{*}\alpha_{3}
\right)
\varphi(\Lambda)^{*}
+ \left( \alpha_{1}\alpha_{2}^{*} -\alpha_{1}^{*}\alpha_{2} \right)
\varphi^{\prime}(\Lambda) \nonumber \\ 
& \ \ = 
\left(
\alpha_{1}\varphi(-\Lambda) +\alpha_{2}\varphi^{\prime}(-\Lambda)
\right)^{*}.
\label{eq:theo-4-2}
\end{eqnarray}
Since $\alpha_{1} \alpha_{2}^{*} \in \mathbb{R}$ by Lemma \ref{lem:3}, 
we have 
\begin{equation}
\alpha_{1} \alpha_{2}^{*} -\alpha_{1}^{*} \alpha_{2} =0.
\label{eq:theo-4-3}
\end{equation} 
Since $\alpha_{1} \alpha_{4}^{*} \in \mathbb{R}$ by Lemma \ref{lem:3} again, 
we have $\alpha_{1}^{*} \alpha_{4} = \alpha_{1} \alpha_{4}^{*}$. 
It follows from this fact and ($\mathcal{A}$1) that 
\begin{equation}
\alpha_{1}^{*} \alpha_{4} - \alpha_{2}^{*} \alpha_{3} = 1.
\label{eq:theo-4-4}
\end{equation} 
Combining Eqs. (\ref{eq:theo-4-2}), (\ref{eq:theo-4-3}), 
and (\ref{eq:theo-4-4}), we can conclude that 
$\varphi(\Lambda)=\alpha_{1}\varphi(-\Lambda)
+\alpha_{2}\varphi^{\prime}(-\Lambda)$. 
In the same way as demonstrated above, we obtain 
$\varphi^{\prime}(\Lambda) =\alpha_{3} \varphi(-\Lambda) + \alpha_{4}\varphi'(-\Lambda)$. 
Thus, $\varphi \in D(H_{\vec{\alpha}})$, that is, 
$H_{\vec{\alpha}}^{*} \subset H_{\vec{\alpha}}$. 
Hence it follows from the two arguments 
that $H_{\vec{\alpha}}$ is self-adjoint, and thus, 
part (\ref{theo:4:1}) is completed. 

Part (\ref{theo:4:2}) directly follows from Lemma \ref{lem:7}. 
\end{proof}
To see the correspondence of 
our boundary condition and Eq. (2.2) of Ref.~\cite{eg}, we show the following lemma: 
\begin{lemma}
\label{lem:eg}
Let $\vec{\alpha}$ be in the class $\mathcal{A}$. 
If $\alpha_{j^{\prime}}\ne0$, then 
$\alpha_{j}\alpha_{j'}^{-1} \in \mathbb{R}$ 
for $j, j^{\prime}=1, 2, 3, 4$.  
\end{lemma}
\begin{proof}
\proof Since $\alpha_{j}\alpha_{j^{\prime}}^{-1}
=\alpha_{j}\alpha_{j^{\prime}}^{*}|\alpha_{j^{\prime}}|^{-2}\in\mathbb{R}$ 
by Lemma \ref{lem:3}, we obtain the desired result. 
\end{proof}
\begin{remark}
\label{rem:tunneling} 
We assume that the wave function $\psi(x)$ is in 
$D(H_{0}^{*})$ for every $\Lambda>0$ so that 
\begin{equation}
\left\{ \begin{array}{l}
\lim_{\Lambda\to 0}\psi(-\Lambda)=\psi(0-), \\ 
\lim_{\Lambda\to 0}\psi(+\Lambda)=\psi(0+), \\
\lim_{\Lambda\to 0}\psi^{\prime}(-\Lambda)=\psi^{\prime}(0-), \\ 
\lim_{\Lambda\to 0}\psi^{\prime}(+\Lambda)=\psi'(0+).
\end{array} \right.
\end{equation}
Let $\vec{\alpha}$, $a$, $b$, and $c$ 
be given as $\vec{\alpha} \in \mathcal{A}$, 
$a, b \in\mathbb{R}$, and $c\in\mathbb{C}$, respectively. 
When $\vec{\alpha}$ with $\alpha_{2}\ne 0$ 
is given arbitrarily, based on Lemma \ref{lem:eg}, 
we set $a, b$, and $c$ as 
\begin{equation}
a:=\alpha_{4}\alpha_{2}^{-1}\in\mathbb{R},\quad  
b:=\alpha_{1}\alpha_{2}^{-1}\in\mathbb{R},\quad 
c:=\, -(\alpha_{2}^{*})^{-1}\in\mathbb{C}.        
\end{equation}
So, we have 
$\alpha_{3}=\, -(c^{*})^{-1}(|c|^{2}-ab)$. 
Conversely, when $a, b \in\mathbb{R}$, 
and $c\in\mathbb{C}$ with $c\ne 0$ 
are given arbitrarily, 
we set $\vec{\alpha}$ as 
\begin{eqnarray}
& \alpha_{1}:=\, -(c^{*})^{-1}b, \quad  
& \alpha_{2}:=\, -(c^{*})^{-1}, \nonumber \\
& \alpha_{3}:=\, (c^{*})^{-1}(|c|^{2}-ab),\quad  
& \alpha_{4}:=\, -(c^{*})^{-1}a.  
\end{eqnarray}
Let us assume that the boundary condition in Theorem \ref{theo:4} 
holds for all $\Lambda>0$. 
Then, our boundary condition in Theorem \ref{theo:4} 
tends to the boundary condition Eq.(2.2) of Ref.~\cite{eg} 
as $\Lambda\to 0$:  
\begin{eqnarray}
\psi^{\prime}(0+)=a\psi(0+)+c\psi(0-), \nonumber \\ 
-\psi^{\prime}(0-)=c^{*}\psi(0+)+b\psi(0-). 
\end{eqnarray}
\end{remark}
As a special case of Theorem \ref{theo:4}, 
we take $\vec{\alpha}$ given in Example \ref{ex:1}. 
That is, we set $-\alpha_{2}=\alpha_{3}=e^{i\theta}$ for 
every $\theta \in \left[\left. 0 , 2\pi \right)\right.$ and $\alpha_{1}=0=\alpha_{4}$.  
For $\alpha:=(\alpha_{2} , \alpha_{3})$ 
we define the action of the Hamiltonian $H_{\alpha}$ by 
$H_{\alpha}:= -d^{2}/dx^{2}$, and the domain $D(H_{\alpha})$ by 
\begin{equation}
D(H_{\alpha}):= 
\left\{ 
\psi\in D(H_{0}^{*})\, |\, 
\psi(\Lambda)=\alpha_{2}\psi^{\prime}(-\Lambda)
\ \textrm{and} \ 
\psi^{\prime}(\Lambda)=\alpha_{3}\psi(-\Lambda)
\right\}. 
\end{equation}
Then, Theorem \ref{theo:4}~(\ref{theo:4:1}) says that 
$H_{\alpha}$ is a self-adjoint extension of $H_{0}$. 
This comes up with a concrete phase factor in the boundary condition as an example. 
Let $\gamma_{{}_{\rightarrow}}$ and $\gamma_{{}_{\leftarrow}}$ be given 
by $\gamma_{{}_{\rightarrow}}:=e^{i\left\{\theta + \sqrt{2}\Lambda + (3\pi/4)\right\}}$ 
and 
$\gamma_{{}_{\leftarrow}}:=e^{i\left\{-\theta + \sqrt{2}\Lambda + (3\pi/4)\right\}}$ 
for arbitrary $\theta$ with $\theta\in \left[\left. 0,2\pi\right)\right.$. 
For $\gamma:=(\gamma_{{}_{\rightarrow}} , \gamma_{{}_{\leftarrow}})$ 
we define the action of the Hamiltonian $H_{\gamma}$ by 
$H_{\gamma}:= -d^{2}/dx^{2}$, 
and give its domain $D(H_{\gamma})$ by the set of all 
wave functions $\psi$ satisfying Eq. (\ref{eq:tunneling}): 
\begin{eqnarray}
D(H_{\gamma}):=
& \left\{  
\psi_{0}+c_{{}_{\mathrm{L}}}(L_{+}+\gamma_{{}_{\rightarrow}}R_{-})
+c_{{}_{\mathrm{R}}}(R_{+}+\gamma_{{}_{\leftarrow}}L_{-}) \,\,\, | \right. \nonumber \\ 
& \left. \ \ \ \ \ \ \ \ \ \ \ \ \ \ \ \ \ \ \ \ \ \ \ \ \ \ \ \ \ \ \ \ \ \ \ \ \ \
\psi_{0}\in D(H_{0}), c_{{}_{\mathrm{L}}}, c_{{}_{\mathrm{R}}} \in \mathbb{C}
\right\}. 
\end{eqnarray}
Then, Theorem \ref{theo:4} says that $H_{\gamma}$ is 
a self-adjoint extension of $H_{0}$, 
and that $H_{\alpha}$ is represented by $H_{\gamma}$. 
Moreover, $H_{\alpha}$ and $H_{\gamma}$ have the one-to-one 
correspondence as in the following theorem:
\begin{theorem} 
\label{theo:2} 
Let $\theta\in [0 , 2\pi)$ 
and $\alpha_{2}, \alpha_{3} 
\in\mathbb{C}$ be given arbitrarily. 
Then, any subspace $D(H_{\gamma})$ 
and any subspace $D(H_{\alpha})$ 
are equal if and only if 
the correspondence: 
\begin{eqnarray}
\alpha_{2}= 
-e^{i\theta}\,\,\,\,\, 
\textrm{and}\,\,\,\,\, 
\alpha_{3}= e^{i\theta}.
\label{eq:correspondence-non-isolated}
\end{eqnarray}
holds. 
\end{theorem}
\begin{proof}
Assume $D(H_{\gamma})=D(H_{\alpha})$. 
Take an arbitrary vector $\psi\in D(H_{\gamma})$. 
It is equivalent to take the vector 
$\psi=\psi_{0}+c_{{}_{\mathrm{L}}}L_{+}+c_{{}_{\mathrm{R}}}R_{+}
+c_{{}_{\mathrm{L}}}\gamma_{{}_{\rightarrow}}R_{-}
+c_{{}_{\mathrm{R}}}\gamma_{{}_{\leftarrow}}L_{-}$ 
for arbitrary $c_{{}_{\mathrm{L}}}, c_{{}_{\mathrm{R}}}\in
\mathbb{C}$ and arbitrary $\psi_{0}\in D(H_{0})$.
By the boundary condition, we have 
\begin{equation}
\left\{ \begin{array}{l}
\gamma_{{}_{\rightarrow}}R_{-}(\Lambda)-e^{-i\pi/4}\alpha_{2}L_{+}(-\Lambda)
=0, \\ 
R_{+}(\Lambda)-e^{i\pi/4}\alpha_{2}\gamma_{{}_{\leftarrow}}L_{-}(-\Lambda)=0,
\end{array} \right.
\label{eq:lem2-1}
\end{equation}
and 
\begin{equation}
\left\{ \begin{array}{l}
e^{-i3\pi/4}\gamma_{{}_{\rightarrow}}R_{-}(\Lambda)
-\alpha_{3}L_{+}(-\Lambda)=0, \\ 
e^{i3\pi/4}R_{+}(\Lambda)-\alpha_{3}\gamma_{{}_{\leftarrow}}L_{-}(-\Lambda)=0. 
\end{array} \right.
\label{eq:lem2-2}
\end{equation}
Using Eqs. (\ref{eq:LR-property}), (\ref{eq:RR^*-property}), 
(\ref{eq:lem2-1}), and (\ref{eq:lem2-2}), 
we obtain 
\begin{equation}
\left\{ \begin{array}{l}
\gamma_{{}_{\rightarrow}}-e^{-i\pi/4}\alpha_{2}e^{i\sqrt{2}\Lambda}=0, \\ 
e^{i\sqrt{2}\Lambda}-e^{i\pi/4}\alpha_{2}\gamma_{{}_{\leftarrow}}=0,
\end{array} \right.
\end{equation}
and 
\begin{equation}
\left\{ \begin{array}{l}
e^{-i3\pi/4}\gamma_{{}_{\rightarrow}}-\alpha_{3}e^{i\sqrt{2}\Lambda}=0, \\ 
e^{i3\pi/4}e^{i\sqrt{2}\Lambda}-\alpha_{3}\gamma_{{}_{\leftarrow}}=0. 
\end{array} \right.
\end{equation}
Eq. (\ref{eq:correspondence-non-isolated}) follows from these four equations.

Conversely, as a corollary of Theorem \ref{theo:4}~(\ref{theo:4:2}), 
Eq. (\ref{eq:correspondence-non-isolated}) implies the equality $D(H_{\gamma})=D(H_{\alpha})$. 
\end{proof}
For $\gamma_{{}_{\rightarrow}}$ and $\gamma_{{}_{\leftarrow}}$ 
determined by $\alpha_{2}=-e^{i\theta}$ and $\alpha_{3}=e^{i\theta}$ 
of Theorem \ref{theo:2}, we define 
two functions $f$ and $g$ by 
\begin{equation}
f := L_{+}+\gamma_{{}_{\rightarrow}}R_{-}\,\,\, 
\textrm{and}\,\,\, 
g := R_{+}+\gamma_{{}_{\leftarrow}}L_{-}, 
\label{eq:fg}
\end{equation}    
respectively. 
We introduce a new inner product $(\,\,\,|\,\,\,)$ by 
$(\varphi|\phi):=
\langle\varphi|\phi\rangle 
+\langle\varphi''|\phi''\rangle$. 
We say that $\varphi$ and $\phi$ are $H_{0}$-\textit{diagonal} 
if $(\varphi|\phi)=0$. 
Then, we obtain the following: 
\begin{proposition}
\label{theo:3}
Fix an arbitrary $\theta$ with $0 \le \theta < 2\pi$. 
Define $\gamma_{{}_{\rightarrow}}$ and $\gamma_{{}_{\leftarrow}}$ 
by $\gamma_{{}_{\rightarrow}}:=e^{i\left\{\theta + \sqrt{2}\Lambda + (3\pi/4)\right\}}$ 
and 
$\gamma_{{}_{\leftarrow}}:=e^{i\left\{-\theta + \sqrt{2}\Lambda + (3\pi/4)\right\}}$. 
Then, the following (\ref{theo:3:1}) -- (\ref{theo:3:3}) hold: 
\begin{enumerate}
\item $f$ and $g$ defined in Eq. (\ref{eq:fg}) are $H_{0}$-diagonal; \label{theo:3:1}
\item $f(\Lambda)=e^{i\left\{\theta +(3\pi/4)\right\}}f(-\Lambda)$ 
and $g(\Lambda)=e^{i\left\{\theta -(3\pi/4)\right\}}g(-\Lambda)$; \label{theo:3:2}
\item for every $\psi \in D(H_{\gamma})$ 
\begin{equation}
\psi=\psi_{0}+c_{{}_{\mathrm{L}}}f+c_{{}_{\mathrm{R}}}g,
\end{equation}
where $\psi_{0}\in D(H_{0})$ and $c_{{}_{\mathrm{L}}}, 
c_{{}_{\mathrm{R}}} \in \mathbb{C}$ are uniquely determined by 
Eq.(\ref{eq:tunneling}), 
and moreover, $\psi_{0}$, $f$, and $g$ are mutually 
$H_{0}$-diagonal. \label{theo:3:3}
\end{enumerate}
\end{proposition}
\begin{proof}
The condition (\ref{theo:3:1}) is an easy application of 
Lemma on page 138 of Ref.~\cite{rs2}. 
Simple calculations lead to the condition (\ref{theo:3:2}). 
The condition (\ref{theo:3:3}) directly follows from Theorem \ref{theo:2}.
\end{proof}
\begin{remark}
\label{rem:around-boundary}
Proposition \ref{theo:3} says that the functions $f$ and $g$ 
play essential roles to determine wave function $\psi$ 
around the junction since $\psi_{0}(x)=0$ 
for $x$ in the neighborhood of the junction.  
\end{remark}
\begin{remark}
Proposition \ref{theo:3} (\ref{theo:3:2}) says that the function $f$ (resp. $g$)
has the standard Ramsauer-Townsend (RT) effect
when $\theta=\pi/4+2 n\pi$ (resp. $-\pi/4+ 2 n\pi$) for
each $n\in\mathbb{Z}$. Thus, the functions $f$ and $g$ have a generalization of
the RT effect.
\end{remark}
\subsection{Phase of the tunneling Schr\"{o}dinger particle and the exact WKB analysis}
In this subsection, we will explain a physical meaning of Proposition~\ref{theo:3} using the model 
of the non-adiabatic transition with the three energy level. 

Proposition~\ref{theo:3} tells us that 
the Schr\"{o}dinger particle leads to the interference by the tunneling effect except for $f = g = 0$ 
since the wavefunctions $f$ and $g$ get the phase factors $3 \pi / 4$ and $- 3 \pi / 4$ 
from the boundary conditions, respectively. As a naive guess, we state the following remark: 
\begin{remark}
By Definition~\ref{def:3} ($\mathcal{S}$1) and ($\mathcal{S}$2), we can rewrite the boundary condition in
Proposition \ref{theo:3}~(\ref{theo:3:2}) as
\begin{equation}
f(\Lambda)=\, -e^{i\left\{
\theta - \pi/4
\right\}}f(-\Lambda)\quad
\textrm{and}\quad
g(\Lambda)=\, -e^{i\left\{
\theta + \pi/4
\right\}}g(-\Lambda).
\end{equation}
We recall that, when we apply the WKB
approximation to the Schr\"{o}dinger particle's
barrier-penetration problem, the phase factor $\pi/4$ appears
because of the connection formulas  (see Ref.~\cite[Sec. 12]{B79}).
It should be noted that we have not yet shown whether there is a relation between
that phase factor $\pi/4$ and our $\pi/4$ yet.
\end{remark}
We discuss a connection between our phase factors and the WKB analysis in the following.
Let us assume the model of the Landau-Zener transition for three 
levels~\cite{Aoki} in the tunneling junction as 
\begin{equation}
	i \frac{d}{dt} \left( \begin{array}{c} \psi_1 \\ \psi_2 \\ \psi_3 \end{array} \right) 
	= \eta \left[ \left( \begin{array}{ccc} b_1 t + a & 0 & 0 \\ 0 & b_2 t & 0 \\ 0 & 0 & b_3 t \end{array} \right) + 
	\frac{1}{\sqrt{\eta}} \left( \begin{array}{ccc} 0 & c_{12} & c_{13} \\ \overline{c_{12}} & 0 & c_{23} \\ \overline{c_{13}} & \overline{c_{23}} & 0 \end{array} \right)
	\right] \left( \begin{array}{c} \psi_1 \\ \psi_2 \\ \psi_3 \end{array} \right), 
	\label{WKB}
\end{equation}
where $a > 0$ is constant and $b_3 > b_2 > b_1 > 0$.
The WKB solution of Eq. (\ref{WKB}) is given by Eq. (2.4) of Ref.~\cite{Aoki}. 
The phase factors, which corresponds to one obtained by Proposition~\ref{theo:3}, appear 
in a particular model with the connection matrix for the WKB solution,  given by by Eq. (2.41) of Ref.~\cite{Aoki} from the exact WKB analysis 
up to the order $\eta^{-1/2}$. The connection matrix is computed through the connection formulas Eqs. (2.27), (2.32), and (2.37) of Ref.~\cite{Aoki}.
Furthermore, Eq.~(\ref{WKB}) can be mapped to the BNR equation~\cite{BNR}. According to Ref.~\cite{BNR}, 
the phase factors is obtained by the turning point of the Stokes and anti-Stokes lines. This situation 
may be experimentally realizable in the systems introduced in Sec.~\ref{section:intro}.

A keen reader may notice a relationship between the energy crossing and the self-adjointness. However, 
in this system, the total energy can be preserved while the energy crossing occurs inside the junction. That is, 
the concept of the self-adjoint extension effectively may lead to the energy crossing inside the 
junction remaining the preservation of the total energy. This might be an example of physical meanings on the self-adjointness.
\section{Conclusion and Discussions} \label{section:conc}
We have considered the phase factor of the one-dimensional 
Schr\"{o}dinger particle with the junction like the connected carbon nanotubes 
and shown that this phase factor depends on the situation of the particle, whether the 
particle goes through the junction or not. Theorem \ref{theo:1} means that 
the phase factor of the non-tunneling Schr\"{o}dinger particle does not appear from the boundary 
condition of the junction. Proposition \ref{theo:3} means that 
the phase factor of the tunneling Schr\"{o}dinger particle appears from the boundary condition 
of the junction. Physically speaking, the wavefunction of tunneling 
Schr\"{o}dinger particle shows the interference pattern. This phase factor corresponds to one 
obtained by the exact WKB analysis in the model of the non-adiabatic transition with the three 
energy levels inside the tunneling junction.

There remain the following problems. First, the geometry of the tunneling junction can be taken as the 
Y-junction scheme~\cite{Tokuno} in the complex plane. The relationship between 
the Y-junction scheme and our obtained phase factor has not been shown. Second,  
our considered model may be also analyzed by the duality of the 
quantum graph~\cite{exner,Cheon}. Finally, the extension to the Dirac particle has not yet been done. 
This situation can be experimentally realized by the helical edge state in the quantum spin Hall system 
by the application to the quantum point contact technique~\cite{Hou,Strom}.
\section*{Acknowledgment} 
One of the authors (MH) thanks Pavel Exner for bringing his unpublished paper~\cite{eg} 
to MH and for useful discussion on it. 
Some of the authors (MH and YS) thank Akio Hosoya for suggesting the correspondence 
to the WKB analysis. YS also thanks Alfred Scharff Goldhaber and Hosho Katsura 
for the useful discussions and Namiko Yamamoto 
for guiding them to the literature on the carbon nanotube~\cite{Carbon}.
MH is supported by JSPS, Grant-in-Aid for Scientific Research (C) 20540171.
YS is also supported by JSPS Research Fellowships for Young Scientists (Grant No. 21008624).

\end{document}